%% file: ms.tex
\documentclass[twocolumn]{aastex631}

\usepackage{multirow}
\usepackage{comment}
\usepackage{amsmath}	% Advanced maths commands

\newcommand{\AR}[1]{} % anand s comments
\newcommand{\niceurl}[1]{\href{#1}{#1}}
\newcommand{\oii}{[\ion{O}{2}]~} 
\newcommand{\oiii}{[\ion{O}{3}]~} 
\newcommand{\zspec}{z_{\rm spec}} % to be used in a $$ environment
\newcommand{\zspecvi}{z_{\rm spec, VI}} % to be used in a $$ environment
\newcommand{\zphot}{z_{\rm phot}} % to be used in a $$ environment
\newcommand{\depthdescr}{5$\sigma$ depths for a 0.45 arcsec radius exponential profile, typical of DESI ELG targets}
\newcommand{\desifootdescr}{The thick black line represents a 14 000 deg$^2$ footprint covered by DESI.}
\newcommand{\efftime}{\texttt{EFFTIME\_SPEC}}
\newcommand{\qavi}{\texttt{QA}_{\rm{VI}}} % to be used in a $$ environment
\newcommand{\coii}{(g-r) + 1.2 \times (r-z)}  % to be used in a $$ environment

\shorttitle{DESI ELG TS}
\shortauthors{Raichoor et al.}

\graphicspath{{plots/}}

\begin{document}

\title{Target Selection and Validation of DESI Emission Line Galaxies}

\correspondingauthor{Anand Raichoor}
\email{araichoor@lbl.gov}
\input{DESI-2021-0104_author_list.tex}

\begin{abstract}
The Dark Energy Spectroscopic Instrument (DESI) will precisely constrain cosmic expansion and the growth of structure by collecting $\sim$40 million extra-galactic redshifts across $\sim$80\% of cosmic history and one third of the sky.
The Emission Line Galaxy (ELG) sample, which will comprise about one-third of all DESI tracers, will be used to probe the Universe over the $0.6 < z < 1.6$ range, which includes the $1.1<z<1.6$ range, expected to provide the tightest constraints.
We present the target selection of the DESI SV1 Survey Validation and Main Survey ELG samples, which relies on the Legacy Surveys imaging.
The Main ELG selection consists of a $g$-band magnitude cut and a $(g-r)$ vs.\ $(r-z)$ color box, while the SV1 selection explores extensions of the Main selection boundaries.
The Main ELG sample is composed of two disjoint subsamples, which have target densities of about 1940 deg$^{-2}$ and 460 deg$^{-2}$, respectively.
We first characterize their photometric properties and density variations across the footprint.
Then we analyze the DESI spectroscopic data obtained since December 2020 during the Survey Validation and the Main Survey up to December 2021.
We establish a preliminary criterion to select reliable redshifts, based on the \oii~flux measurement, and assess its performance.
Using that criterion, we are able to present the spectroscopic efficiency of the Main ELG selection, along with its redshift distribution.
We thus demonstrate that the the main selection with higher target density sample should provide more than 400 deg$^{-2}$ reliable redshifts in both the $0.6<z<1.1$ and the $1.1<z<1.6$ ranges.
\end{abstract}


%=======================================================
% Section : Introduction
%=======================================================
\section{Introduction} \label{sec:intro}

% probes, facilities
Since the observation of the acceleration of the expansion of the Universe \citep{riess98a,perlmutter99a}, the cosmology community has focused its efforts to gather the data providing the more precise possible observational constraints.
Several cosmological probes are used in order to perform independent measurements with different systematics (see \citealt{weinberg13a} for a review), the most established methods being
	Type Ia supernovae and baryonic acoustic oscillations (BAO) to constrain the geometry of the Universe,
	and weak-lensing, galaxy clusters, and redshift-space distortions (RSD) to constrain the growth of structures.
To reach that goal, dedicated facilities will survey large fractions of the sky with high-quality imaging (e.g., DES: \citealt{des-collaboration05a}, HSC: \citealt{aihara18a}, \textit{Euclid}: \citealt{laureijs11a}, LSST: \citealt{ivezic19a}) and/or massive spectroscopy (e.g., 2dF: \citealt{colless03a}, 6dF: \citealt{jones09a}, BOSS: \citealt{dawson13a}, WiggleZ: \citealt{drinkwater10a}, eBOSS: \citealt{dawson16a}, DESI, \textit{Euclid}, PFS: \citealt{takada14a}).

% massive spectro. surveys
Massive spectroscopic surveys probe the large-scale structures (LSS) of the matter distribution, by measuring the spectroscopic redshift ($\zspec$) of a vast number of galaxies over large areas and different epochs.
One strength of this approach is that the very same dataset allows one to constrain at the same time the geometry of the Universe with the BAO scale \citep{eisenstein98a} and the growth of structures with the RSD \citep{kaiser87a} method.
The SDSS experiment \citep{york00a} has been a pioneer of such surveys, with the co-first BAO measurement \citep{eisenstein05a}.
\citet{alam21a} summarize and analyze twenty years of SDSS spectroscopic observations of about two million $\zspec$ over $0 < z < 5$ and 10,000 deg$^2$, which led to state-of-the-art constraints on the Hubble constant ($H_0 = 68.18 \pm 0.79$ km.s$^{-1}$.Mpc$^{-1}$) and the $\sigma_8$ parameter normalizing the growth of structures ($\sigma_8 = 0.85 \pm 0.03$).

% desi
The DESI experiment \citep{levi13a,desi-collaboration16a,desi-collaboration16b} will pursue this effort and increase the number of observed $\zspec$ by an order of magnitude, with about 40 million extra-galactic $\zspec$ over 14,000 deg$^2$.
DESI will follow the same approach as SDSS, and use an optimized tracer for each targeted redshift range.
About 13 million bright galaxy sample (BGS) will cover the $0.05 < z < 0.4$ range, 
about 8 million luminous red galaxies (LRG) will cover the $0.4 < z < 1.1$ range,
about 16 million emission line galaxies (ELG) will cover the $0.6 < z < 1.6$ range,
and lastly about 3 million quasars (QSO) will cover the $z > 0.9$ range, used as tracers in the $0.9 < z < 2.1$ range, and using Ly-$\alpha$ forests as a probe of the intergalactic medium at $z > 2.1$.
Additionally, DESI will also observe about 10 million stars from the Milky Way Survey (MWS).
The BGS and MWS programs will be observed in `bright' time, i.e. when the Moon is up, whereas the other tracers (LRG, ELG, QSO) will be observed in `dark' time, i.e. when the Moon is down.

% desi elg
This paper is dedicated to the DESI ELG sample, which is composed of star-forming galaxies.
The principle of the ELG sample is to take advantage of the two following facts:
	(1) the Universe star-formation rate density peaks at $z \sim 1$-2 \citep[e.g.,][]{madau14a}, thus star-forming galaxies were very common at that epoch;
	(2) the ELG $\zspec$ can be reliably measured in a rather short amount of observation time, as it only requires a significant detection of the emission lines in the spectrum, with no need to significantly detect the continuum; in particular, the \oii doublet $\lambda \lambda$ 3726,29 \AA~offers an unambiguous signature of the $\zspec$ (see for instance \citealt{moustakas06a} for the link between the \oii line-strength and the star formation).
Some reference intensive spectroscopic surveys sampled that ELG population at $z \sim 1$-2 over few square degrees (e.g., VVDS: \citealt{le-fevre13a}, or DEEP2: \citealt{newman13a}), paving the way for their use in spectroscopic cosmological experiments.
For the above reasons, the ELG tracer is a key tracer of this decade massive spectroscopic surveys (e.g., DESI, \textit{Euclid}, PFS), and will constitute about one-third of DESI spectra.
The DESI ELG sample will probe the Universe over the $0.6 < z < 1.6$ range, and in particular over the $1.1 < z < 1.6$ range, which will bring the tightest of the DESI cosmological constraints.
It will be the very first to densely sample this redshift range, providing faint targets not extensively explored by any survey.
For instance, the eBOSS ELG sample \citep{raichoor17a,raichoor21a} has a target density about ten times smaller and targets about one magnitude brighter than the DESI ELG sample.

% desi elg requirements
In that respect, the DESI ELG sample will be the first of its kind, and thus provides several challenges.
First, the target density needs to be high, about 2400 deg$^{-2}$ because ELG targets will be assigned fibers after the LRG and QSO targets, hence the selection must provide enough targets so that a target can be reached by each fiber most of the time.
This requires to select rather faint targets; however, another constraint is the requirement that the number of $\zspec$ measurement failures in a typical DESI exposure (15 minutes in nominal conditions) remain a reasonable fraction of observed spectra.
For that purpose, a large enough fraction of the targets needs to have sufficient \oii flux to secure a reliable $\zspec$ measurement.
A quantified requirement is that the DESI ELG target sample provides at least 400 deg$^{-2}$ reliable $\zspec$ in both the $0.6 < z < 1.1$ and $1.1 < z < 1.6$ ranges, as Fisher forecasts show that this is sufficient to reach the required cosmological precision of the DESI experiment \citep{desi-collaboration22b}.
Lastly, as for other tracers, the DESI ELG sample must have a fraction of catastrophic $\zspec$ measurements ('catastrophics`) as low as possible (of the order of one percent), the LSS analysis being very sensitive to catastrophics $\zspec$ \AR{reference?}.

% sv, main surveys
To meet these requirements, the DESI experiment conducted a Survey Validation program (December 2020 to May 2021) before starting the actual Main Survey in May 2021.
The first part of the Survey Validation program, hereafter called SV1 (December 2020 to March 2021)  consisted in deep observations of extended target selections for all tracers.
Those SV1 data have been used to fine-tune the Main Survey target selections.
The second part of the Survey Validation program was the One-Percent Survey, hereafter called One-Percent (April -- May 2021), where target selections close or equal to the Main Survey ones were observed at a very high completeness.

% other kpy1 papers
This paper is part of a series of papers presenting the DESI target selections and their characterization.
\citet{desi-collaboration22b} present an overview of the DESI spectroscopic observations and the tracers used by those papers, and \citet{myers22a} present how those target selections are implemented in DESI.
\citet{lan22a} and \citet{alexander22a} present the construction of spectroscopic truth tables based from visual inspection (VI) for the galaxies (BGS, LRG, ELG) and the QSO targets, respectively.
The MWS sample is presented in \citet{cooper22a}, 
the BGS sample is presented in \citet{hahn22a},
the LRG sample is presented in \citet{zhou22a},  
the ELG sample is presented in this paper, and
the QSO sample is presented in \citet{chaussidon22a}.
Those five target selection papers present the DESI final samples, and superseed the preliminary target selections presented in \citet{allende-prieto20a}, \citet{ruiz-macias20a}, \citet{zhou20a}, \citet{raichoor20a}, and \citet{yeche20a}.

% paper plan
This paper is structured as follows.
Section~\ref{sec:photdata} presents the imaging, the footprints, and  the photometry used to select the ELG targets.
We then present the Main Survey ELG target selection in Section~\ref{sec:maints}, the SV1 ELG target selection in Section~\ref{sec:svts}, and the Main Survey ELG sample photometric properties in Section~\ref{sec:mainphot}.
Section~\ref{sec:specdata} introduces the DESI spectroscopic data (SV1, One-Percent, and Main survey observations up to December 2021), used in Section~\ref{sec:mainspec} to analyze the spectroscopic properties of the Main Survey ELG sample.
We conclude in Section~\ref{sec:concl}.

% conventions
All magnitudes are in the AB system \citep{oke83a}, and corrected for Galactic extinction using the \citet{schlegel98a} maps.
All displayed sky maps use the \texttt{Healpix} scheme \citep{gorski05a} with a resolution of 0.21 deg$^2$ (\texttt{nside} = 128), but the computation in Section~\ref{sec:photsyst} uses a finer resolution of 0.05 deg$^2$ (\texttt{nside} = 256).\\

%=======================================================
% Section : Imaging and photometry
%=======================================================
\section{Imaging, footprints, and photometry} \label{sec:photdata}

The DESI ELG targets are selected from the $grz$-photometry of the DR9 release of the Legacy Imaging Surveys\footnote{\niceurl{https://www.legacysurvey.org/dr9}} \citep[LS-DR9;][]{schlegel22a}.
This release covers about 19,700 deg$^2$ in the optical $grz$-bands -- complemented with the Wide-field Infrared Survey Explorer (\textit{WISE}) near-infrared data \citep{meisner21a}.
We present here a brief description of the optical imaging and the photometry, focusing on the part relevant for the ELG target selection, and refer the reader to \citet{schlegel22a} for more details.

% section: imaging
\subsection{Imaging} \label{sec:imaging}
The optical $grz$-imaging comes from several observing programs.
The northern part of the North Galactic Cap (NGC) comes from two programs: the Beijing-Arizona Sky Survey \citep[BASS,][]{zou17a} provides the $g$- and $r$-bands observed with the 90Prime camera on the Bok 2.3 m telescope; and the Mayall z-band Legacy Survey (MzLS) provides the $z$-band observed with the Mosaic-3 camera on the 4 m Mayall telescope at Kitt Peak National Observatory (KPNO).
The southern part of the NGC and the South Galactic Cap (SGC) mostly come from two programs, the Dark Energy Camera Legacy Survey \citep[DECaLS,][]{dey19a} and the Dark Energy Survey \citep[DES,][]{des-collaboration05a}; both use the Dark Energy Camera \citep[DECam][]{flaugher15a} on the 4 m Blanco telescope at the Cerro Tololo Inter-American Observatory (CTIO).

We note that the DECaLS, BASS, and MzLS surveys followed a dynamic observing strategy to achieve as much as possible a uniform depth across the footprints.
In particular, the considered depths account for the Galactic extinction, i.e. the imaging is deeper in regions with high Galactic extinction, so that the target selection should be less sensitive to the Galactic dust map (see Section 6.2 of \citealt{dey19a}).
Nevertheless, because of the capping of the individual imaging exposure time, this strategy cannot be applied for the $g$-band imaging for $\rm{E(B-V)} \gtrsim 0.15$, as the $g$-band has the largest extinction factor\footnote{The A/E(B-V) coefficients are  3.214, 2.165, and 1.211 for the $g$-, $r$-, and $z$-band, respectively; see \niceurl{https://www.legacysurvey.org/dr9/catalogs/\#galactic-extinction-coefficients}}.
The DES survey did not follow such a strategy, but as it is fairly deep, the effect of the Galactic extinction on the imaging depth is less critical for the ELG targets.

\begin{figure}[!h]
	\begin{center}
		\begin{tabular}{c}
			\includegraphics[width=0.95\columnwidth]{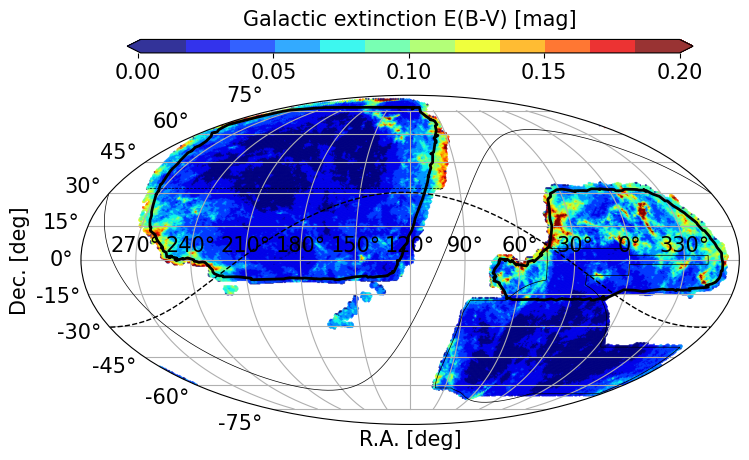}\\
			\includegraphics[width=0.95\columnwidth]{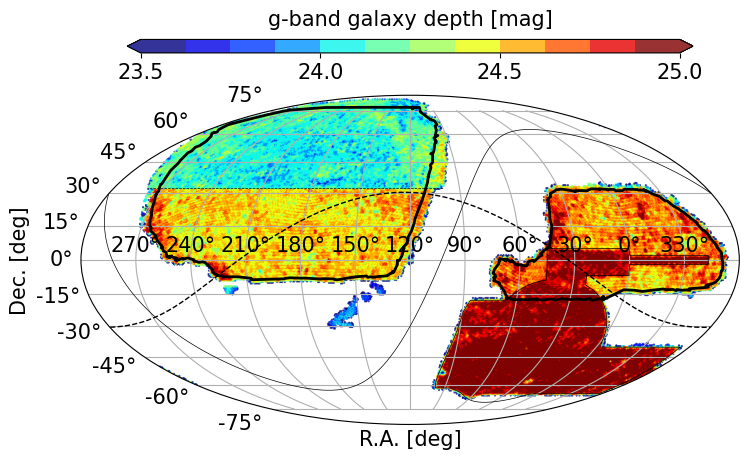}\\
			\includegraphics[width=0.95\columnwidth]{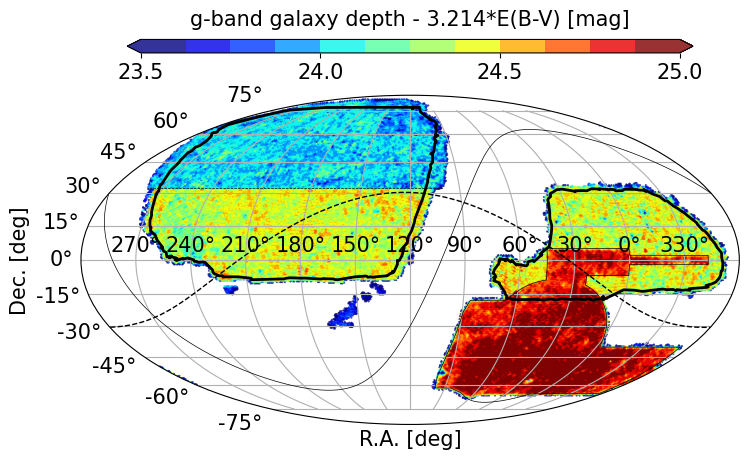}\\
		\end{tabular}
		\caption{
			Sky maps of the Galactic dust extinction (top), the imaging $g$-band depth (middle), and the extinction-corrected $g$-band depth (bottom) for the imaging data used to select the DESI ELG targets.
			Depths are \depthdescr.
			\desifootdescr{}
			The Galactic plane is displayed as a solid line and the Sagittarius plane as a dashed line.
		}
		\label{fig:ebvdepth}
	\end{center}
\end{figure}

Figure~\ref{fig:ebvdepth} illustrates this approach for the $g$-band where the extinction factor is the largest.
In regions of high extinction (top panel), the $g$-band depth (middle panel) is larger, which results in a rather homogeneous extinction-corrected depth map in each of the three footprints of the imaging program (bottom panel).

% figure: sky ebv, depth

% section: footprints
\subsection{Footprints} \label{sec:footprints}
\begin{figure}[!h]
	\begin{center}
		\includegraphics[width=0.95\columnwidth]{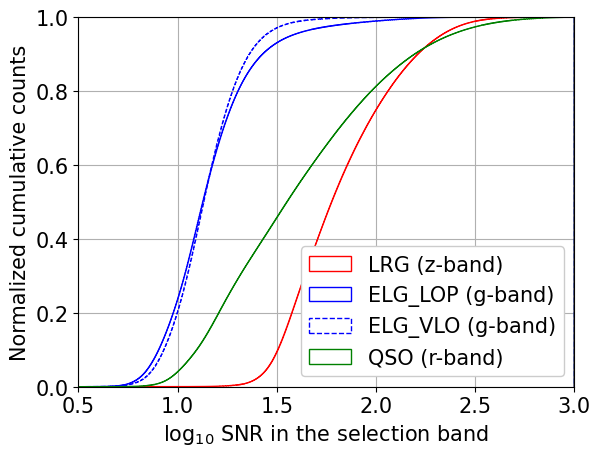}
		\caption{
			Cumulative distribution of the logarithm of the photometric SNR (see text) in the selection band for the DESI Main Survey dark tracers over the nominal DESI footprint.
			For the ELG targets, the two Main Survey selections are shown.
		}
		\label{fig:photsnr}
	\end{center}
\end{figure}

As a result of the different programs providing the imaging, different parts of the footprint have different imaging depths, which matters for the ELG target selection, as those tracers are faint in imaging.
To illustrate this point, Figure~\ref{fig:photsnr} displays the normalized cumulative distributions of the signal-to-noise ratio (SNR) in the selection band\footnote{For instance $\rm SNR = \texttt{flux\_g} \times \sqrt{\texttt{flux\_ivar\_g}}$ for the ELG targets.} over the nominal DESI footprint for the three dark tracers of  DESI.
The ELG targets have a typical SNR of 13 in the imaging, whereas the QSO and LRG targets have a typical SNR of 36 and 60, respectively, and thus are less sensitive to the depth variations across the footprints.

For that reason, we define three footprints, which will be analyzed separately in this paper:
	the North, corresponding to the Dec $>$ 32.375$^\circ$ part of the NGC, covered by BASS and MzLS;
	the South-DECaLS, corresponding to the non-DES SGC part and the Dec. $<$ 32.375$^\circ$ part of the NGC;
	the South-DES, corresponding to the DES imaging in the SGC.
Those footprints can be visualized in the depth maps in Figure~\ref{fig:ebvdepth}, where the North is displayed in blue-green, the South-DECaLS in yellow-orange, and the South-DES in red.
Table~\ref{tab:imagdens} reports the approximate areas and imaging depths per footprint.
One notices that the North footprint is about 0.5 mag shallower in the $g$- and $r$-bands than the South-DECaLS footprint, and the South-DES footprint is about 0.5-1.0 mag deeper than the South-DECaLS footprint in all three $grz$-bands.
As ELG targets are faint, those depth differences impact the target selection, in terms of detected objects and contamination, as it will be seen further in the paper.\\

% table: imaging properties per footprints
\begin{table*}
	\centering
	\begin{tabular}{lccccccc}
		\hline
		\hline
		Footprint & LS-DR9 area & DESI area & $g$-depth & $r$-depth & $z$-depth & ELG\_LOP density & ELG\_VLO density\\
		& $[$deg$^2]$ & $[$deg$^2]$ & $[$AB mag$]$ & $[$AB mag$]$ & $[$AB mag$]$ & $[$deg$^{-2}]$ & $[$deg$^{-2}]$\\
		\hline
		% clean
		North		 & 5100 & 4400 & 24.1 & 23.5 & 23.0 & 1930 & 410\\
		South-DECaLS & 9500 & 8500 & 24.5 & 23.9 & 23.0 &  1950 & 490\\
		South-DES	 & 5100 & 1100 & 24.9 & 24.7 & 23.5 & 1900 & 480\\
	        \hline
    \end{tabular}
	\caption{
		Imaging properties and ELG target density per footprint for each of the two Main ELG samples (ELG\_LOP and ELG\_VLO).
		Areas are approximate.
		Depths are 5-$\sigma$ depths for a 0.45 arcsec radius exponential profile, typical of DESI ELGs.
	}
	\label{tab:imagdens}
\end{table*}

% section: photometry
\subsection{Photometry} \label{sec:photometry}
The overall data reduction and photometry is performed with the \texttt{legacypipe}\footnote{\niceurl{https://github.com/legacysurvey/legacypipe}} pipeline.
The LS-DR9 images are astrometrically calibrated with \textit{Gaia} DR2 \citep{gaia-collaboration18a} and photometrically calibrated with Pan-STARRS 1 \citep{chambers16a}, using color terms to place the photometry on the same system as the LS-DR9 one \AR{add sentence about ubercal for dec$<$-30}.
The photometry is performed with the \texttt{Tractor} software \citep{lang16a,lang22a}.
Source detection is done on stacked images, then all measurements are based on individual exposures. 
Each source is modeled with an analytic profile (point-source, exponential with fixed parameters, exponential, de Vaucouleurs or S\'{e}rsic) and a model image is generated for each exposure.
Increasingly more complex profiles are allowed for sources detected with higher SNR.
The source properties (position, shape, flux) are measured through a likelihood optimization ($\chi^2$ minimization) of the set of model images covering the considered region.

% total and fiber flux
Based on the best-fit properties, \texttt{Tractor} provides the total flux of each source, as well as its 'fiber flux`, which corresponds to the predicted flux within a fiber of diameter 1.5 arcsec -- the size of a DESI fiber -- for 1 arcsec Gaussian seeing.
Those fiber fluxes hence predict the typical amount of light that a DESI fiber would see.

% figure: photsnr

%=======================================================
% Section : Main Target Selection
%=======================================================
\section{Main Survey Target Selection} \label{sec:maints}

In this section, we present the DESI Main Survey ELG target selection.
The selection cuts are detailed in Table~\ref{tab:maincuts} and illustrated in Figure~\ref{fig:maingrz}.
The extended selection explored in the DESI SV1 program, which was used to finalized those Main Survey cuts, is presented in Section~\ref{sec:svts}.

The DESI ELG sample is the first of its kind, with no previous significant reference sample observed so far.
For instance, the VVDS or DEEP2 observations covered few square degrees; the WiggleZ \citep{drinkwater10a} or the eBOSS/ELG \citep{raichoor17a, raichoor21a} surveys observed about one thousand square degrees, but their ELG samples were more than one magnitude brighter, have a five-to-ten times lower density (about 200-400 deg$^{-2}$) and only extends to $z < 1.1$.
The first proposed DESI ELG selection was based on a simple $g-r$ vs. $r-z$ selection \citep{desi-collaboration16a}, though it could not be spectroscopically tested at that time.
\citet{karim20a} explored that selection, along with more advanced ones, with dedicated spectroscopical observations.
This pilot program demonstrated that all tested selections had overall similar performances.

Based on that analysis, for the sake of simplicity and robustness, we chose simple color-color cuts in the  $g-r$ vs. $r-z$ diagram to select the DESI ELG targets.
We describe hereafter the chosen cuts.

% elg_lop, elg_vlo
\subsection{ELG\_LOP and ELG\_VLO subsamples} \label{sec:mainsubs}
As mentioned in Section \ref{sec:intro}, the goal of the DESI ELG sample is to provide cosmological constraints over the $0.6 < z < 1.6$ range, with favoring as much as possible the $1.1 < z < 1.6$ range, where other DESI samples are the least dense.
To do so, the ELG sample is split in two disjoint samples, ELG\_LOP at $\sim$1940 deg$^{-2}$ and ELG\_VLO at $\sim$460 deg$^{-2}$\footnote{The area to compute those densities does not account for the $\sim$1\% area removed with the angular masking described in Section \ref{sec:qualcuts}.}.
We remind that the DESI observations use priorities to assign fibers to targets \citep{raichoor22b}.
ELG\_LOP has higher priority in the fiber assignment and favors the $1.1 < z < 1.6$ range, whereas ELG\_VLO has lower priority and favors the $0.6 < z < 1.1$ range.
With that fiber assignment configuration, cosmological Fisher forecasts demonstrate that the ELG sample fulfills the expected performance \citep{desi-collaboration22b}.

The names of those two samples, ELG\_LOP and ELG\_VLO, are names assigned to targeting bits by \texttt{desitarget}, the target selection pipeline \citep{myers22a}, and indicate their ELG priority state in the fiber assignment ('low` and `very-low').

% elg_hip
\subsection{ELG\_HIP subsample} \label{sec:mainhip}
For the dark tiles, the tracers in order of decreasing fiber assignment priorities are: QSO, LRG, ELG\_LOP, and ELG\_VLO.
This results in very-high fiber assignment rates for the QSO and LRG targets, but lower ones for the ELG\_LOP targets, and even lower ones for the ELG\_VLO targets.
In order to still have a significant number of observed pairs of ELG and LRG targets, a third ELG sample is defined, ELG\_HIP, which is a ten percent random subsampling of the ELG\_LOP and ELG\_VLO samples, with the same fiber assignment priority  that the one of the LRG targets.
This provides more information about the small-scale cross-correlation between the ELG targets and higher priority targets.
Without this extra sample, the small-scale effects of fiber collisions, together with the preference for always observing the higher priority objects, would significantly increase the noise for cosmological analyzes cross-correlating ELG targets with LRG targets \citep[see e.g.,][for the weights computation method]{bianchi18a, mohammad20a}.

Similarly to ELG\_LOP and ELG\_VLO, the ELG\_HIP name is assigned to targeting bits by \texttt{desitarget}, and indicates the ELG priority state in the fiber assignment ('high`).
As this ELG\_HIP sample is a random subsample of the ELG\_LOP and ELG\_VLO samples, we do not discuss it further in this paper.

% section: main cuts
\subsection{Main Survey selection cuts} \label{sec:maincuts}

\begin{figure}[!h]
	\begin{center}
		\includegraphics[width=0.95\columnwidth]{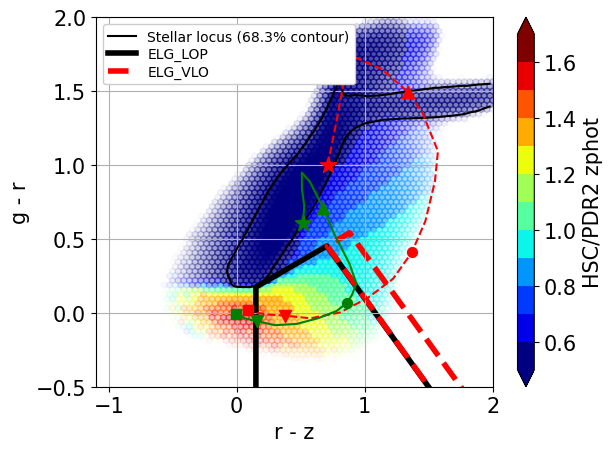}
		\caption{
			Main ELG target selection cuts in the $g-r$ vs. $r-z$ diagram.
			The ELG\_LOP and the ELG\_VLO selections are displayed as solid black lines and dashed red lines, respectively.
			The background symbols describe a $g_{\rm fib} < 24.1$ sample; the color-coding indicates the mean HSC/DR2 $\zphot$ measurements and the transparency scales with the logarithm of the density.
			The thin black contour indicates the stellar locus.
			Tracks show stellar evolutions for a highly star-forming (green solid line) and a less star-forming (red dashed line) galaxy observed at $\zspec$ 2.0, 1.6, 1.1, 0.6, 0.1 (square, down triangle, circle, up triangle, star, respectively), see text for more details.}
		\label{fig:maingrz}
	\end{center}
\end{figure}

The Main Survey ELG selection cuts are detailed in Table \ref{tab:maincuts} and illustrated in Figure \ref{fig:maingrz}.
Those are of three kinds:
	(1) quality cuts to ensure that the photometry is reliable;
	(2) a cut in the $g$-band fiber magnitude;
	(3) a selection box in the $g-r$ vs. $r-z$ diagram.

% figure: main grz

We underline that the cuts are the same in the North and in the South-DECaLS/DES footprints, even though the photometric systems are slightly different.
This choice was motivated by several reasons:
	the exact color transformation between the two systems is not trivial, as it depends on the considered object (e.g., star, blue or red galaxies);
	the North has different imaging systematics than the South, in particular in the $g$-band depth;
	with the partial sampling of the Survey Validation program it was not possible to define a secure tuning of the selection cuts, which would provide a similar ELG redshift distribution in the three footprints, as the redshift distribution has non-trivial dependencies on the imaging and foreground variations.
For the sake of simplicity, we thus keep the same cuts in the three footprints.

% main cuts: quality cuts
\subsubsection{Quality cuts} \label{sec:qualcuts}
Quality cuts are designed to select sources with reliable photometric measurements.
For computation reasons, the \texttt{legacypipe} pipeline processes the sky in $0.25^\circ \times 0.25^\circ$ bricks, which slightly overlap.
The \texttt{brick\_primary} cut requests the object not to be in the LS-DR9 overlap between two bricks.
We further require that there is at least one observation in each of the three $grz$-bands, and that the measured flux as a positive SNR in all three bands (i.e., a positive flux and a non-null inverse variance).
And we apply a minimal angular masking, to reject regions around very bright stars ($\textit{Gaia} \; \rm G < 13$) and large galaxies \citep{moustakas22a} or globular clusters.
We emphasize that this masking is purposely minimal, and common to all three DESI dark tracers.
Further \textit{a posteriori} masking will be applied on the spectroscopic data with the analyze thereof, as for instance we know from the target density variations only that the ELG target selection has spurious targets around moderately bright stars ($13 < \textit{Gaia} \; \rm G < 16$).

% main cuts: g-magnitude cut
\subsubsection{$g$-band magnitude cut}
We then make a selection on the $g$-band magnitude, which is motivated by the fact that the \oii flux correlates best with the bluest band flux \citep{comparat16a}; this ensures that the selection favors \oii emitters.
We discard bright objects, which are unlikely to be at $z > 0.6$; as those represent a marginal fraction of the ELG sample, it does not matter if we cut on the fiber or total magnitude.
The faint cut in the $g$-band fiber magnitude is tuned to reach the desired densities of about 1940 deg$^{-2}$ for ELG\_LOP and 460 deg$^{-2}$ for ELG\_VLO.
A cut on the fiber magnitudes was favored over a cut in total magnitudes, because the latter provides more $\zspec$ failures due to galaxies with not enough flux in the DESI fiber.

% main cuts: grz box
\subsubsection{$g-r$ vs. $r-z$ selection}
The ELG\_LOP and ELG\_VLO samples rely on simple $g-r$ vs. $r-z$ cuts.
The primary motivation of the cuts is the redshift selection, as illustrated in Figure~\ref{fig:maingrz}.
That figure displays the density of $g_{\rm fib}<24.1$ objects in the LS-DR9 catalogs, where the color-coding indicates the mean photometric redshift ($\zphot$) from the HSC/DR2 data release \citep{aihara19a,tanaka18a}.
Those $\zphot$ measurements are of exquisite quality for our magnitude and redshift range of interest, thanks to the depth and wavelength coverage of the HSC data; \citet{karim20a} already illustrated this point with previous HSC data and we further illustrate in Appendix~\ref{app:hsczphot} how they compare with the DESI ELG $\zspec$.

The slanted cut with a positive slope ($g - r < 0.5 \times (r - z) + 0.1$) in common to ELG\_LOP and ELG\_VLO selections rejects stars and galaxies at $z < 0.6$.

The ELG\_LOP $r-z>0.15$ cut rejects $z > 1.6$ galaxies, for which the \oii doublet is outside the DESI spectrograh coverage so that no reliable $\zspec$ can be expected.
The ELG\_LOP slanted cut with a negative slope ($g - r < -1.2 \times (r - z) + 1.3$) optimizes the fraction of $1.1 < z < 1.6$ targets with high \oii flux, as this is the goal for this sample.
At first order, the redshift is driving this cut, as shown by the HSC $\zphot$ measurements.
At second order, favoring the \oii emitters pushes this cut to the blue, as illustrated by the stellar evolution tracks on Figure~\ref{fig:maingrz}.
Those tracks show two simple \citet{bruzual03a} evolution models of galaxies computed with \texttt{EzGal} \citep{mancone12a}.
The two galaxy models are formed at $z = 3$ with simple exponentially declining star formation histories (i.e., with a star formation rate proportional to $e^{-\rm{age} / \tau}$).
One is moderately star-forming ($\tau = 1$ Gyr, red dashed line), the other one is more star-forming ($\tau = 5$ Gyr,  green solid line); the symbols illustrate where such galaxies would be in the $g-r$ vs. $r-z$ diagram for different observation redshifts: as expected, at a fixed redshift, galaxies with bluer colors are more star-forming.
For instance one sees at $z = 1.1$ (circle) that the most star-forming model is about 0.5 magnitude bluer in $r-z$ than the other one.

The ELG\_VLO slanted cuts with a negative slope ($(g - r > -1.2 \times (r - z) + 1.3)$ and $(g - r < -1.2 \times (r - z) + 1.6)$) are an extension of the ELG\_LOP selection towards redder colors, hence lower redshifts.
The reddest cut is driven to remove $z < 0.6$ galaxies.
We remind that the ELG\_VLO sample is disjoint from the ELG\_LOP sample.

% main cuts: densities
\subsubsection{Target density}
The cuts described above provide an ELG\_LOP sample of about 1940 deg$^{-2}$ and an ELG\_VLO sample of about 460 deg$^{-2}$.
Because of the different imaging properties -- in particular depths -- of the three North, South-DECaLS, South-DES footprints, the actual average density over each footprint is slightly different, as reported in the last two columns of Table~\ref{tab:imagdens}: from 1900 deg$^{-2}$ to 1950 deg$^{-2}$ for the ELG\_LOP sample and from 410 deg$^{-2}$ to 490 deg$^{-2}$ for the ELG\_VLO sample.\\

% Table: main cuts
\begin{table*}
	\centering
	\begin{tabular}{lccl}
		\hline
		\hline
		Sample & Density & Cuts & Comment\\
		\hline
		% clean
		\multirow{4}{*}{Clean} & \multirow{4}{*}{-} & \texttt{brick\_primary} = True & Unique object\\
		& & $\texttt{nobs\_\{grz\}} > 0$ & Observed in the $grz$-bands\\
		& & $\texttt{flux\_\{grz\}} \times \sqrt{\texttt{flux\_ivar\_\{grz\}}} > 0$ & Positive SNR in the $grz$-bands\\
		& & $(\texttt{maskbits} \; \& \; 2^1) = 0$, $(\texttt{maskbits} \; \& \; 2^{12}) = 0$, $(\texttt{maskbits} \; \& \; 2^{13}) = 0$ & Not close to bright star/galaxy\\
		\hline
		% ELG_LOP
		\multirow{5}{*}{ELG\_LOP} & \multirow{5}{*}{$\sim$1940 deg$^{-2}$} & Clean & Clean sample\\
		& & $(g > 20)$ and $(g_{\rm fib} < 24.1)$ & Magnitude cut\\
		& & $0.15 < r - z$ & $r - z$ cut\\
		& & $g - r < 0.5 \times (r - z) + 0.1$ &Star/low-z cut\\
		& & $g - r < -1.2 \times (r - z) + 1.3$ & Redshift/\oii cut\\
        \hline
		% ELG_VLO
		\multirow{5}{*}{ELG\_VLO} & \multirow{5}{*}{$\sim$460 deg$^{-2}$} & Clean & Clean sample\\
		& & $(g > 20)$ and $(g_{\rm fib} < 24.1)$ & Magnitude cut\\
		& & $0.15 < r - z$ & $r - z$ cut\\
		& & $g - r < 0.5 \times (r - z) + 0.1$ &Star/low-z cut\\
		& & $(g - r > -1.2 \times (r - z) + 1.3)$ and $(g - r < -1.2 \times (r - z) + 1.6)$ & Redshift/\oii cut\\
        \hline
    \end{tabular}
	\caption{
		Main Survey target selection cuts.
		The cuts are the same for the North and South-DECaLS/DES regions.
		We use the following definitions:
        $\{grz\} = 22.5 - 2.5 \cdot \rm{log}_{10}(\texttt{flux\_\{grz\}} / \texttt{mw\_transmission\_\{grz\}})$, $g_{\rm fib} = 22.5 - 2.5 \cdot \rm{log}_{10}(\texttt{fiberflux\_g} / \texttt{mw\_transmission\_g})$.
	    The \texttt{brick\_primary}, \texttt{nobs\_\{grz\}}, \texttt{flux\_\{grz\}}, \texttt{fiberflux\_g}, \texttt{flux\_ivar\_\{grz\}}, \texttt{mw\_transmission\_\{grz\}}, \texttt{maskbits} columns are described here: \niceurl{https://www.legacysurvey.org/dr9/catalogs/}.
	}
	\label{tab:maincuts}
\end{table*}

%=======================================================
% Section : Extended (SV1) Target selection
%=======================================================
\section{SV1 Target selection} \label{sec:svts}

In this section, we describe the DESI SV1 target selection, which expands the Main Survey selections described in Section~\ref{sec:maints}.
The SV1 selection cuts are detailed in Appendix~\ref{app:svcuts} and Table~\ref{tab:svcuts}, and illustrated in Figure~\ref{fig:svgrz}.

% section: motivation
\subsection{Motivations} \label{sec:svmotivations}
The SV1 ELG sample was designed to provide informations to finalize the Main Survey ELG selections.
The only existing magnitude-limited, spectroscopic reference samples probing the desired DESI ELG magnitudes are limited to a few square degrees (e.g., DEEP2: \citealt{newman13a}, VVDS: \citealt{le-fevre13a}).
Besides, DESI being a new instrument, its ability to measure reliable $\zspec$ for targets as faint as ELG ones needs to be thoroughly tested.
For those two reasons, the DESI SV1 ELG sample is exploring a rather large photometric space, with a target density of about 7000 deg$^{-2}$.

% section: sv cuts
\subsection{SV1 selection cuts} \label{sec:svcuts}
We hereafter detail the philosophy of the DESI SV1 ELG selection cuts reported in Table~\ref{tab:svcuts} and illustrated in Figure~\ref{fig:svgrz}.

% sv cuts: blue, red, lowz
\subsubsection{$g-r$ vs. $r-z$ extensions}

% sv cuts: blue
The first explored extensions relax the Main cuts in the $g-r$ vs. $r-z$ diagram, as illustrated in the top panel of Figure~\ref{fig:svgrz}.
The cuts are generously extended towards bluer $r-z$ colors, with a $g-r < 0.2$ cut to securely remove low-redshift galaxies and stars.
According to HSC $\zphot$ measurements, that region should include a significant fraction of redshifts in the range $1.1 < z < 1.6$ and has been extremely poorly explored so far.
While this region is very valuable for DESI, with $1.1 < z < 1.6$ ELG targets, it is also costly because any $z > 1.6$ target would not provide a reliable $\zspec$, as the \oii doublet is outside of the DESI spectrograph coverage.

% sv cuts: lowz
The cuts are also slightly extended towards the low-redshift galaxies and stellar locus (positive slope cut).
Existing spectroscopic data and HSC $\zphot$ consistently show that there is a sharp transition, with a density of $z<0.6$ objects quickly rising when going to redder $g-r$ colors.
As DESI is expected to provide reliable $\zspec$ for most of those, there only is a marginal need to explore that region.

% sv cuts: red
Lastly, the cuts are extended towards redder $r-z$ colors, to cover the eBOSS/ELG selection region.
From HSC $\zphot$ and eBOSS/ELG $\zspec$ measurements, we know that this region mostly hosts $z<1.1$ galaxies.
This extension is motivated by the early desire to have an overlap with the eBOSS/ELG sample, and to secure a fallback Main selection in the unlikely case the DESI instrument were performing far worse than expected.

% sv cuts: faint (sliding cut)
\subsubsection{Faint extensions (sliding cut)}
\begin{figure}[!h]
	\begin{center}
		\begin{tabular}{c}
			\includegraphics[width=0.95\columnwidth]{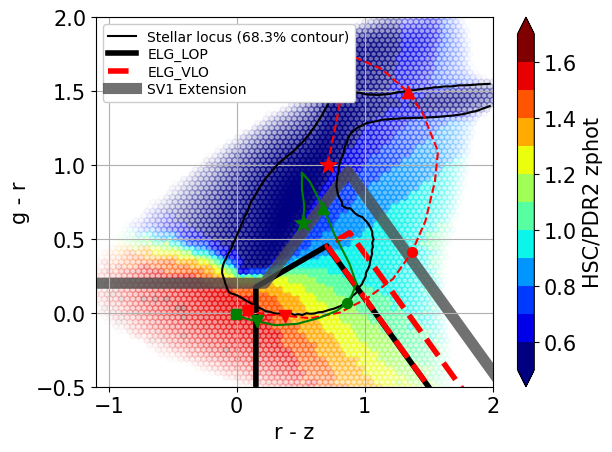}\\ 
			\\[1pt]
			\includegraphics[width=0.95\columnwidth]{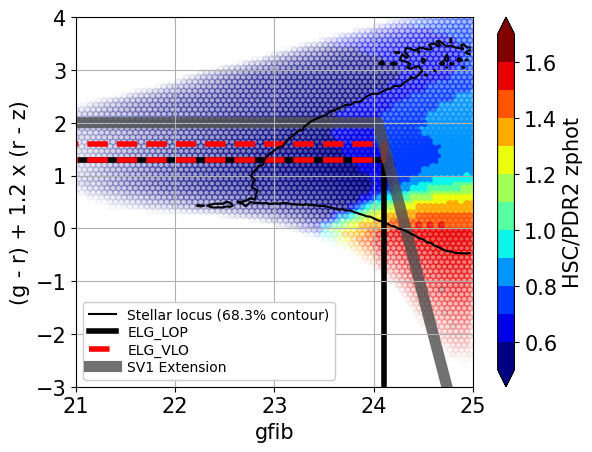}\\
		\end{tabular}
		\caption{
			SV1 ELG target selection cuts in the $g-r$ vs. $r-z$ diagram (top) and $\coii$ vs $g_{\rm fib}$ diagram (bottom).
			The ELG\_LOP and the ELG\_VLO selections are displayed as solid black lines and dashed red lines, respectively.
			The background symbols describe a $g_{\rm fib} < 25.0$ sample; the color-coding indicates the mean HSC/DR2 $\zphot$ measurements and the transparency scales with the logarithm of the density.
			Contour and tracks are as in Figure~\ref{fig:maingrz}.
		}
		\label{fig:svgrz}
	\end{center}
\end{figure}

An important extension explores the faint end of the target selection to test the ability of the DESI instrument to provide a reliable $\zspec$ there.
Provided that the target density significantly increases when going fainter, this extension is restricted to blue objects - the most interesting ones for ELGs - to prevent the SV sample to be overwhelmed by faint objects.
To do so, we adopted a sliding cut in the $g-r$ vs. $r-z$ diagram, as illustrated in the bottom panel of the Figure~\ref{fig:svgrz}.
The sliding cut uses the $\coii$ color, which broadly scales as the \oii flux.
On the red $r-z$ side, this cut restricts the sample to bright objects, as faint objects there are expected to have a marginal \oii flux, and thus are unlikely to provide a reliable $\zspec$.
On the blue $r-z$ side, this cut explores targets fainter by few tenths of magnitudes, which are expected to have a significant \oii flux, and hence would provide a reliable $\zspec$.

% sv cuts: gtot vs. gfib mag selection
\subsubsection{$g_{\rm tot}$ and $g_{\rm fib}$ extensions}
Finally, all the above cuts are applied on samples restricted in $g_{\rm tot}$, the total $g$-band magnitude or in $g_{\rm fib}$, the fiber $g$-band magnitude.
While a $g_{\rm tot}$-limited sample corresponds to a better defined galaxy population, it could contain a significant fraction of targets with too small a flux in the DESI fibers to provide a reliable $\zspec$.
A $g_{\rm fib}$-limited sample has the advantage of being more homogeneous and complete in terms of reliable $\zspec$.\\

% figure: sv grz and coii

%=======================================================
% Section : Photometric properties of the Main sample
%=======================================================
\section{Photometric properties of the Main sample} \label{sec:mainphot}
This Section presents a preliminary discussion about the Main Survey ELG sample density fluctuations across the footprint, which are driven by variations of both the LS-DR9 imaging properties and of the astrophysical foreground maps (e.g., Galactic stellar density and dust extinction).
The ELG target magnitudes being close to the imaging depth, this sample is more sensitive to those imaging/foreground variations than other DESI dark tracers, and will likely require significant dedicated work to account for those in data analysis, which needs to beforehand remove such dependencies prior to perform a cosmological analysis.

Besides, the final LSS ELG sample will be restricted to objects with a reliable redshift in the $0.6 < z < 1.6$ range.
Both ELG target density fluctuations and redshift efficiency variations with spectroscopic observing conditions will have to be corrected to produce reliable cosmological results.

This is why we hereafter restrict to simple diagnoses, in order to illustrate the overall properties of the Main Survey ELG sample.

% ts main: magnitude distributions
\subsection{Magnitude distributions}
\begin{figure*}[!h]
	\begin{center}
		\includegraphics[width=0.95\textwidth]{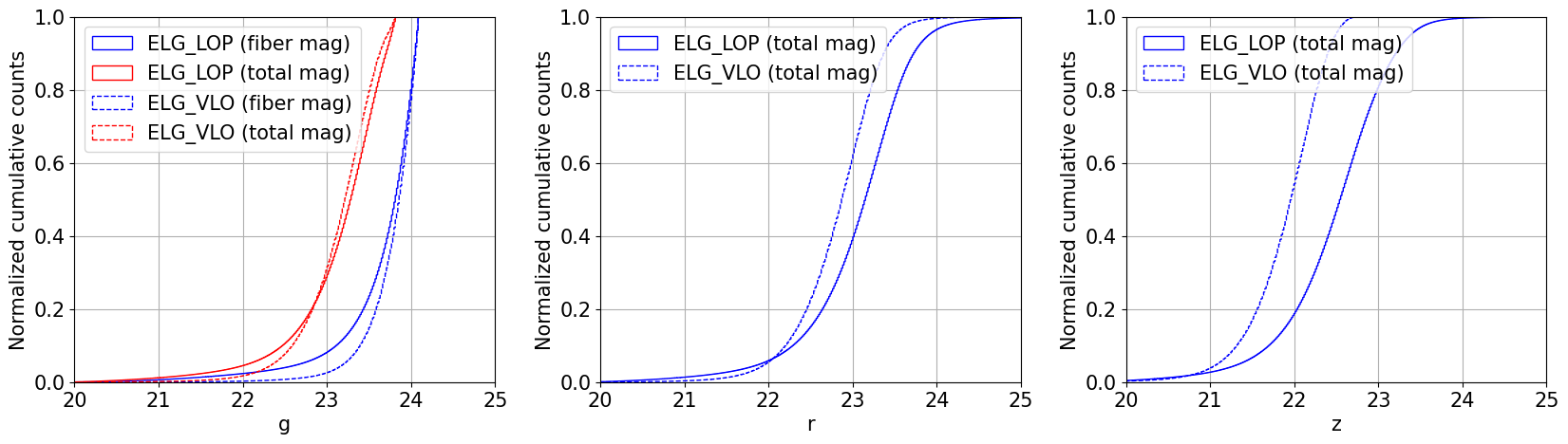}\\ 
		\caption{
			ELG\_LOP (solid lines) and ELG\_VLO (dashed) targets total magnitude normalized cumulative distributions in the $g$-band (left), $r$-band (middle), and $z$-band (right).
			For the $g$-band, we also display the fiber magnitude distributions (red lines).
		}
		\label{fig:mag}
	\end{center}
\end{figure*}

Figure~\ref{fig:mag} displays the ELG\_LOP (solide lines) and ELG\_VLO (dashed lines) normalized cumulative distributions of target magnitudes.
For the $g$-band, we both present the fiber magnitude, used for the selection, and the total magnitude.
For the $r$- and $z$-bands, we only present the total magnitude, as those come at play through colors only.
Both the ELG\_LOP and ELG\_VLO selections are selected with a $g_{\rm fib}<24.1$ cut, hence the two selections present very similar $g$-band magnitude distributions.
However, the different location of the selection boxes in the $g-r$ vs. $r-z$ diagram implies different magnitude distributions in the $r$- and $z$-bands, with the ELG\_LOP targets being 0.2 mag fainter than the ELG\_VLO targets in the $r$-band, and 0.5 mag fainter in the $z$-band.
We thus expect the ELG\_LOP selection to have a stronger dependency with the imaging $z$-band depth.

% figure: magnitude distributions

% ts main: density map
\subsection{Density maps}
\begin{figure*}[!h]
	\begin{center}
		\begin{tabular}{lr}
			\includegraphics[width=1.05\columnwidth]{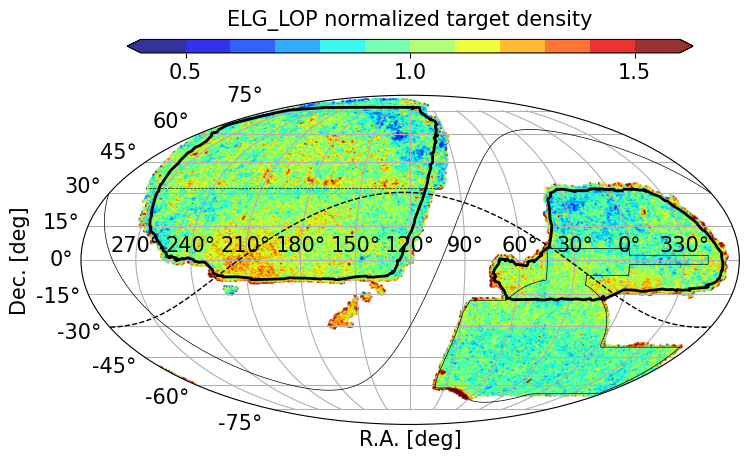} &
			\includegraphics[width=1.05\columnwidth]{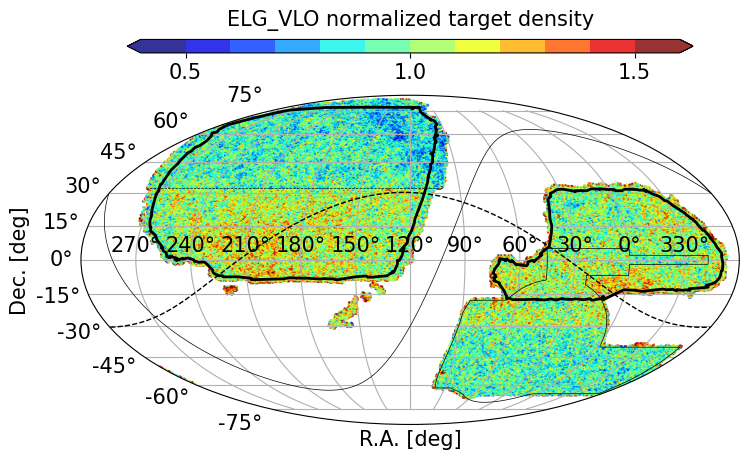}\\
		\end{tabular}
		\caption{
			Main Survey ELG\_LOP (left) and ELG\_VLO (right) sample density sky map.
			The density is divided by the overall average value (1940 deg$^{-2}$ for the ELG\_LOP sample and 460 deg$^{-2}$ for the ELG\_VLO sample), so as to display the fractional difference to the average.
			\desifootdescr{}
			The Galactic plane is displayed as a solid line and the Sagittarius plane as a dashed line.
		}
		\label{fig:sky}
	\end{center}
\end{figure*}

Figure~\ref{fig:sky} displays the density fluctuations of ELG\_LOP (top panel) and ELG\_VLO (bottom panel) targets across the whole LS-DR9 footprint.
A 14 000 deg$^2$ footprint covered by DESI is indicated with thick black contours.
Several features are visible, in particular for the ELG\_LOP sample; we comment the most noticeable ones.

% figure: sky density

\begin{figure}[!h]
	\begin{center}
		\includegraphics[width=0.95\columnwidth]{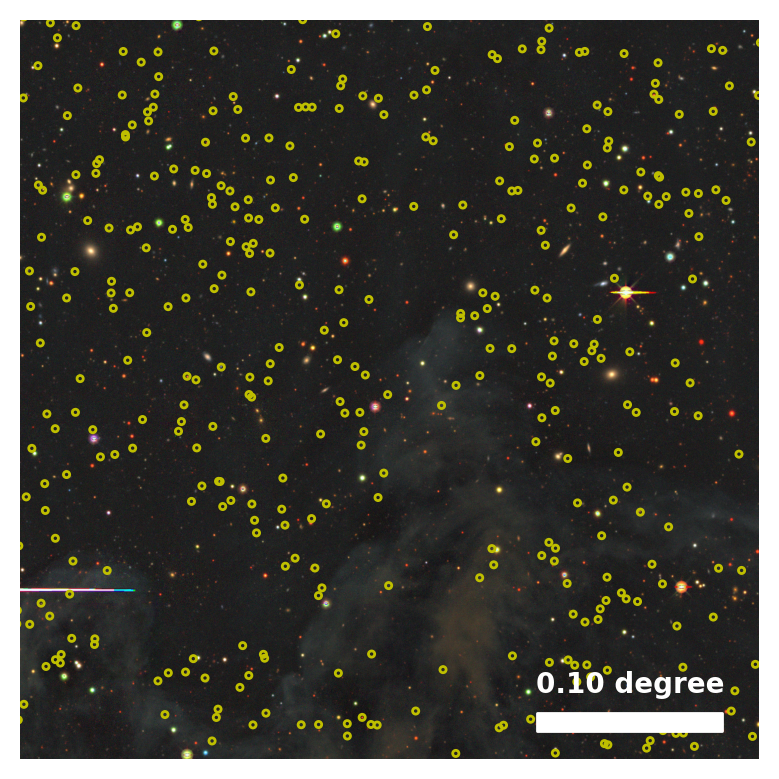}
		\caption{Example of a region with Galactic dust clouds causing extreme extinction variations at small scales, which are imprinted in the ELG\_LOP target sample (yellow circles).
		The cutout is centered at (R.A, Dec.) = (31.40$^\circ$, 20.66$^\circ$) and is 0.4 degree-wide.
		}
		\label{fig:ebvcutout}
	\end{center}
\end{figure}

% ebv
The blue regions at $\rm(R.A., Dec.) \sim (130^\circ, 60^\circ)$ or $\rm(R.A., Dec.) \sim (30^\circ, 20^\circ)$ are under-densities due to a region of high Galactic dust, with many small-scale structures, as illustrated in Figure~\ref{fig:ebvcutout}.
A possible interpretation could be that,  even though the imaging is deeper in dusty regions of the footprint, the extinction effect is only partially corrected in those regions (see Section~\ref{sec:imaging}), in particular high variations of the dust extinction at small scale cannot be handled at the imaging level.
However, we note that some high-extinction regions can show an excess of ELG\_LOP targets, as for instance at $\rm(R.A., Dec.) \sim (345^\circ, 20^\circ)$.
Proper explanations of those effects likely require a detailed analysis of the interplay of the target selection with the dust extinction, the imaging depth, and the behavior of the \texttt{Tractor} source detection and fitting in those regions.
Approaches like \texttt{Obiwan} \citep{kong20a}, which injects fake sources in the imaging itself and then runs \texttt{Tractor}, may bring key information for such issue.

% figure: ebv cutout

% sag. stream
The ELG\_LOP sample seems to have an overdensity along the Sagittarius Stream, displayed as a dashed line in Figure~\ref{fig:sky}.
The Sagittarius Stream has a stellar population bluer than the Galactic population \AR{ref?} and could in principle add contaminants to the ELG\_LOP selection.
As of now it is not clear whether the ELG\_LOP overdensity is due to that Sagittarius Stream stars, or if it is just concomitant: a detailed analysis of the spectroscopic observations will be required to clarify this issue.

% north overdensities
The overdensities in the North at $(\rm{R.A, Dec.}) \sim (180^\circ, 40^\circ)$ or $(210^\circ, 40^\circ)$ correspond to regions of shallower extinction-corrected $g$-band imaging (see bottom plot of Figure~\ref{fig:ebvdepth}).

% des dec = -30 + lmc + ngc -15<dec<30
Lastly, we comment three other features noticeable on those maps, even though they are well outside the DESI footprint, and hence not relevant for the DESI observations.
For both selections, the density becomes slightly smaller below the $\rm{Dec.} = -30^\circ$ latitude in the DES region.
This is due to a known shift of approximately 0.02 mag in the $z$-band, where the calibration method transitions from Pan-STARRS1 to Ubercal \citep{padmanabhan08a,schlegel22a} \AR{check consistency with DR9 paper}.
The large ELG\_LOP overdensity at $\rm{(R.A., Dec.)} \sim (80^\circ, -60^\circ)$ at very South edge of the DES region is contamination from the Large Magellanic Cloud, which adds a high density of blue stars in that region.
And the ELG\_LOP overdensity at $\rm{(R.A., Dec.)} \sim (150^\circ, -20^\circ)$ is due to the much shallower imaging there (see Figure~\ref{fig:ebvdepth}).

% ts main: photometric systematics
\subsection{Photometric systematics} \label{sec:photsyst}
In this Section, we present how the ELG target selection depends on the imaging and foreground properties.
As already stated, ELG targets have magnitudes close to the imaging depth, which makes the sample sensitive to fluctuations in the imaging and foreground maps.

We consider here the simplest set of maps.
For the foreground, those encompass the \textit{Gaia} stellar density and the Galactic dust extinction (E(B-V) parameter); besides, we also consider the projected distance to the Sagittarius Stream, as it has been seen in the previous Section that it could be a relevant quantity to consider.
For the imaging, in each of the three $grz$-bands, we consider the seeing ('PSF size`) and the 'galaxy depth` corrected for dust extinction.
We use dust-extinction corrected depths as the target selection relies on dereddened magnitudes.

\begin{figure*}
	\begin{center}
		\includegraphics[width=0.95\textwidth]{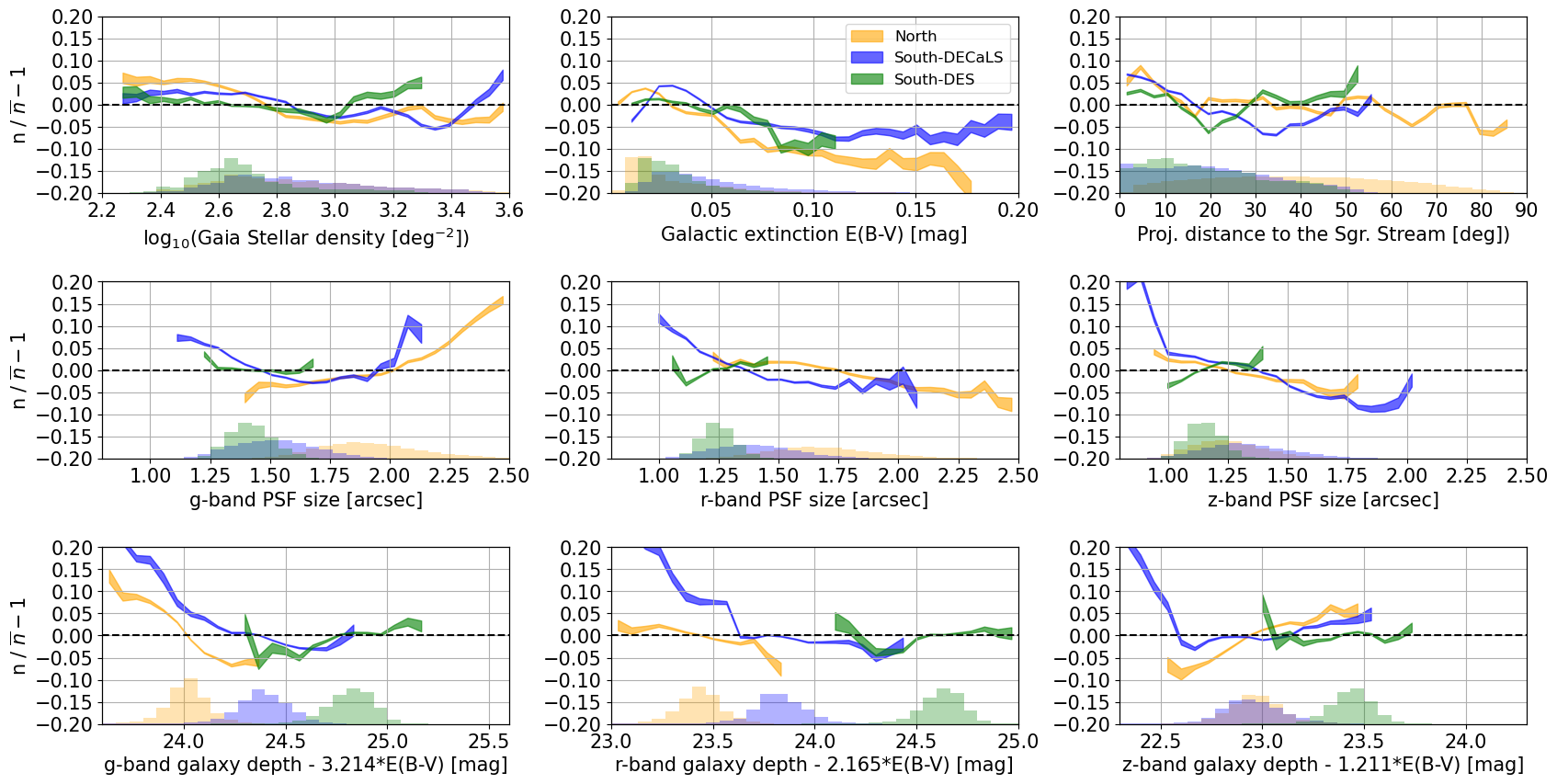}
		\caption{
			Main Survey ELG\_LOP target density variations with foregrounds (top) and imaging seeing (middle) and depth (bottom).
			The South-DES $\rm{Dec.} < -30^\circ$ region has been excluded.
			We consider bins with at least 100 healpix pixels.
		}
		\label{fig:elglopsyst}
	\end{center}
\end{figure*}

\begin{figure*}
	\begin{center}
		\includegraphics[width=0.95\textwidth]{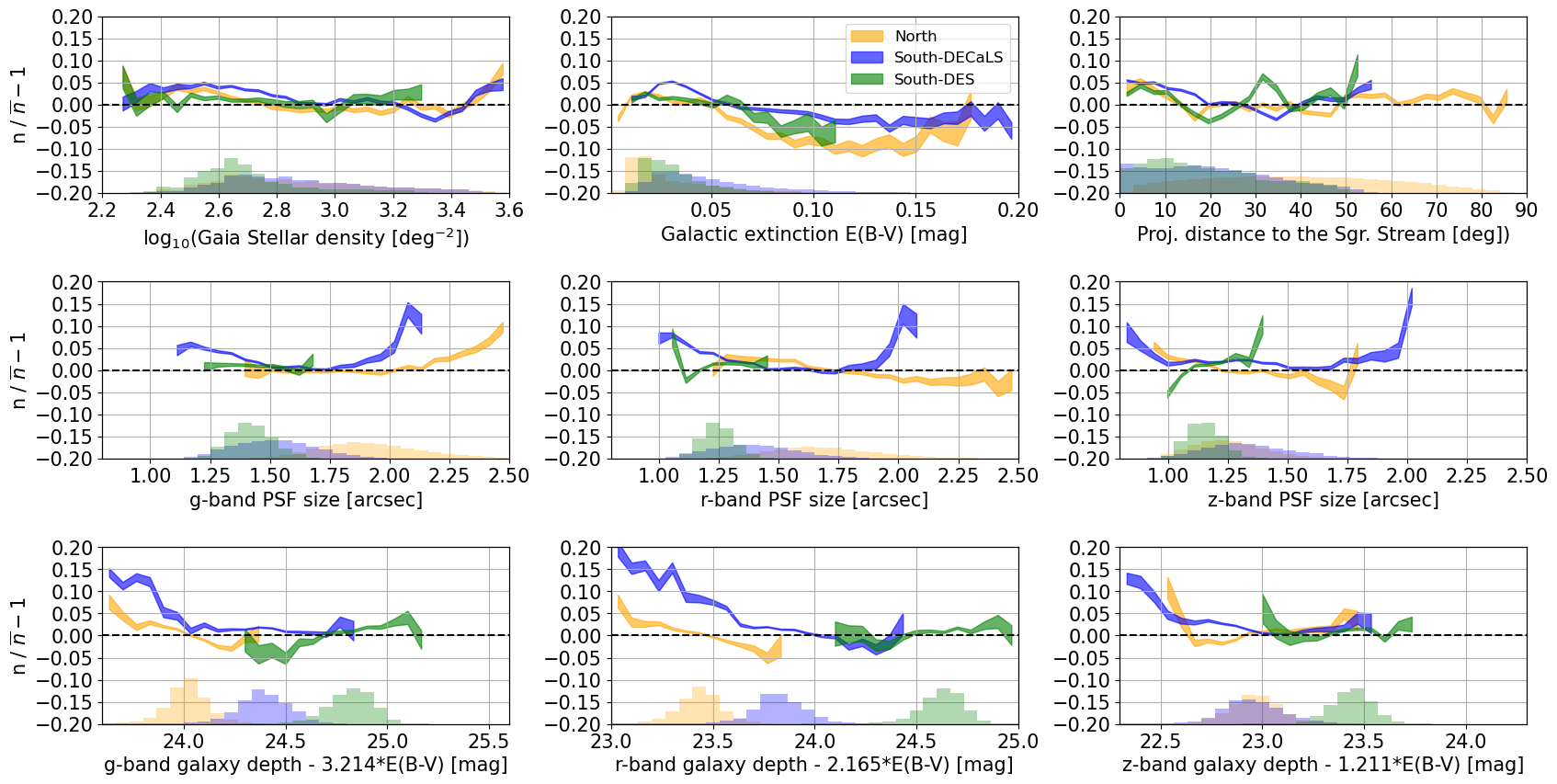}
		\caption{
			Main Survey ELG\_VLO target density variations with foregrounds (top) and imaging seeing (middle) and depth (bottom).
			The South-DES $\rm{Dec.} < -30^\circ$ region has been excluded.
			We consider bins with at least 100 healpix pixels.
		}
		\label{fig:elgvlosyst}
	\end{center}
\end{figure*}

Figures~\ref{fig:elglopsyst} and~\ref{fig:elgvlosyst} show how the ELG\_LOP and ELG\_VLO target selection densities vary with those foreground and imaging maps, for each of the North, South-DECaLS, and South-DES footprints.
We note that we discard here the South-DES $\rm{Dec.} < -30^\circ$ region, because of the small photometric calibration issue mentioned in the previous section.
Those figures are based on 0.05 deg$^2$ \texttt{Healpix} pixels ($\texttt{nside} = 256$).
For each footprint, the density variations are normalized to the average density over the footprint.
The different properties of each footprint are clearly visible in this plot, and illustrate the need to analyze those separately.

In general, both selections display similar trends, the ELG\_LOP sample showing stronger trends, as expected from the fact that it contains fainter objects.

% figure: elg_lop syst

% figure: elg_vlo syst

% syst: g-depth
The most significant dependency is that with the $g$-band depth, which shows two behaviors.
For the North and the South-DECaLS footprints, the density decreases with increasing depth, whereas for the South-DES footprint, it increases with increasing depth.
A possible explanation could be the following: for shallow imaging, the trend would be driven by contamination from stars and $z < 0.6$ galaxies due to scattering in the $g-r$ color, which makes them move inside the selection box.
The scattering also affects $z > 0.6$ galaxies, making them go outside the selection box.
However the density of such $z>0.6$ galaxies is much smaller than that of the $z<0.6$ galaxies (see Figure~\ref{fig:maingrz}).
The net effect would be an increase in the number of selected targets.
The strength of this effect decreases as the depth increases (the color scattering decreasing), and the impact on density eventually vanishes.
Another effect could come at play with deeper imaging, namely the increase in the number of detected sources in the imaging: that second effect could explain the trend seen in the South-DES region.

% syst: ebv
There is a decrease of the target density with the Galactic extinction for $\rm{E(B-V)} > 0.05$ mag, the trend being stronger in the North than in the South-DECaLS and South-DES footprints.
This could be explained by two facts: the imaging depth does not fully correct for the Galactic extinction (Section~\ref{sec:imaging}) and the regions with high extinction are embedded in small-scale structures that cannot be correctly accounted for in the imaging strategy (Figure~\ref{fig:ebvcutout}).

% syst: sgr stream
Interestingly, both selections lead to increased density close to the Sagittarius Stream, which could be explained by contamination from stars from the stream.

% ts main: sensitivity to zeropoint
\subsection{Sensitivity to photometric zero-point uncertainties}
We estimate the sensitivity of the Main Survey ELG target selection to $\sigma_{\rm zp}$, the imaging photometric zero-point uncertainties, using the same approach as in \citet{myers15a} and \citet{raichoor17a}.
Results are reported in Table~\ref{tab:zp}.

In each of the $g$-, $r$-, and $z$-band, one at a time, we add $\pm0.01$ mag to the photometry and re-run the target selection algorithm to estimate $\delta N_{0.01} = \frac{|\Delta N|}{N}$, the fractional change in the target density due to this magnitude shift.
We find consistent $\delta N_{0.01}$ values across the footprints.
The ELG\_LOP selection has $\delta N_{0.01} \sim 0.05, 0.03, 0.01$ in the $g$-, $r$-, and $z$-band, respectively.
The ELG\_VLO selection has $\delta N_{0.01} \sim 0.04, 0.05, 0.04$ in the $g$-, $r$-, and $z$-band, respectively.
We notice that the selections have different sensitivities in the $z$-band, ELG\_VLO being more sensitive.

The expected rms variation in the number density due to shifts of the imaging zero-point is then estimated to be $\frac{\delta N_{0.01}}{0.01} \times \sigma_{\rm zp}$.
The LS-DR9 has $\sigma_{\rm zp}$ values of 0.003 mag in the $g$- and $r$-bands, and of 0.006 mag in the $z$-band \citep{schlegel22a}. \AR{to be confirmed!}
If we assume Gaussian errors for the zero-points, 95 percent of the footprint lie within a $\pm 2 \sigma_{\rm zp}$ of the expected zero-point in any photometric band, meaning that 95 percent of the footprint has a variation in target density lower than $4 \times \sigma_{zp} \times \frac{\delta N_{0.01}}{0.01}$.
The resulting fluctuations for each photometric band are given in Table \ref{tab:zp}.
Both selections have density fluctuations of 1-6 percent in all cases, except for the ELG\_VLO sample in the $z$-band, where the density fluctuations is about 8-9 percent.
That level of fluctuation is reasonable, and should be able to be addressed with the weighting scheme in the LSS analysis.\\

% table: zp uncertainty
\begin{table*}
	\centering
	\begin{tabular}{lcc|cc|cc}
		\hline
		\hline
		Band & Footprint & $\sigma_{\rm{zp}}$ &  \multicolumn{2}{c|}{ELG\_LOP} & \multicolumn{2}{c}{ELG\_VLO}\\
		 &  &  & $\delta N_{0.01}$ & Fluctuations over & $\delta N_{0.01}$ & Fluctuations over\\
		 &  &  & & 95 percent of the area & & 95 percent of the area\\
		 &  & [mag]  & & [percent] & & [percent]\\
		\hline
		\multirow{3}{*}{$g$} & North & \multirow{3}{*}{0.003} & 0.052 & 6.2 & 0.038 & 4.6\\
		 & South-DECaLS & & 0.055 & 6.5 & 0.038 & 4.6\\
		 & South-DES & & 0.052 & 6.2 & 0.040 & 4.8\\
		\hline
		\multirow{3}{*}{$r$} & North & \multirow{3}{*}{0.003} & 0.030 & 3.6 & 0.046 & 5.5\\
		 & South-DECaLS & & 0.036 & 4.3 & 0.045 & 5.4\\
		 & South-DES & & 0.032 & 3.8 & 0.051 & 6.1\\
		\hline
		\multirow{3}{*}{$z$} & North & \multirow{3}{*}{0.006} & 0.004 & 1.0 & 0.035 & 8.4\\
		 & South-DECaLS & & 0.005 & 1.2 & 0.032 & 7.7\\
		 & South-DES & & 0.004 & 0.9 & 0.037 & 9.0\\
	        \hline
    \end{tabular}
	\caption{
		Sensitivity of the Main ELG target selection to the imaging photometric zeropoint uncertainties.
		Column 3 is the imaging photometric zeropoint uncertainties ($\sigma_{\rm zp}$).
		Columns 4 and 6 are the fractional changes in target density due to a $\pm$0.01 mag shift in the zeropoint ($\delta N_{0.01}$).
		Columns 5 and 7 are the expected fluctuations in the number of selected targets over 95 percent of the footprint.
	}
	\label{tab:zp}
\end{table*}

%=======================================================
% Section : Spectroscopic data
%=======================================================
\section{Spectroscopic data} \label{sec:specdata}

\AR{check all numbers with sv overview paper and VI paper!}

We now present preliminary results from the DESI spectroscopic observations of this ELG sample.
Those observations include three phases of the DESI experiment: the SV1, the One-Percent, and the Main surveys.
This Section introduces those observations, along with the reduction of the data, which will be used in Section~\ref{sec:mainspec} to perform the analysis.

The SV1 and One-Percent data presented below will be part of the Survey Validation data released in the DESI Early Data Release \citep{desi-collaboration23a}.

% section: desi instrument
\subsection{The DESI instrument}
The DESI instrument, described in details in \citet{desi-collaboration16b} and \citet{desi-collaboration22a}, is a multi-spectroscopic instrument mounted at the prime focus of the 4m Mayall Telescope at Kitt Peak, Arizona.
The focal plane covers a field of view of about 8 deg$^2$ \citep{miller22a} and is equipped with 5,000 fiber positioners \citep{silber22a} distributed in ten 'petals`.
For each of the 500 fibers of a given petal, the light is dispersed by one of the ten three-arm spectrographs ('B`: 360 nm to 600 nm; 'R`: 560 nm to 780 nm; 'Z`: 740 nm to 990 nm).
The resolving power ($R = \lambda / \Delta\lambda$) increases with the wavelength, from $\sim$2000 at the shortest wavelengths to nearly $\sim$5500 at the longest ones.
The wavelength coverage and the resolving power were designed to ensure that the instrument could measure and resolve the ELG \oii $\lambda \lambda$ 3726,29 \AA~doublet in the $0.6 < z < 1.6$ range.

% section: observations
\subsection{Observations} \label{sec:specobs}
\begin{figure}
	\begin{center}
		\includegraphics[width=0.95\columnwidth]{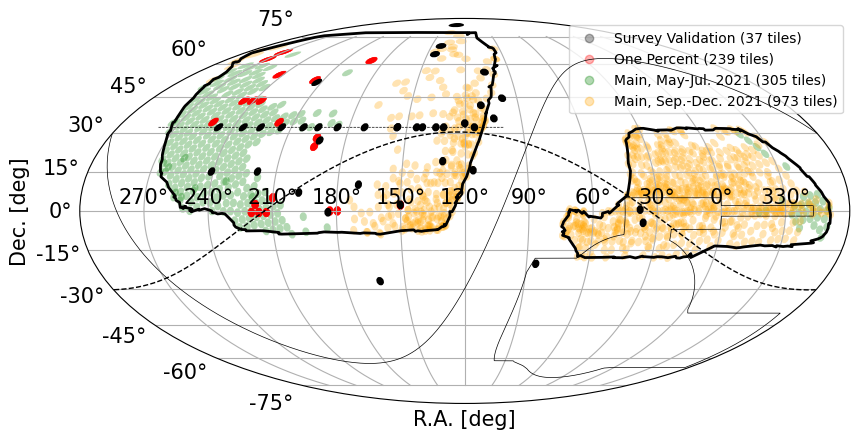}
		\caption{
			Sky distribution of the DESI observed dark tiles used in this paper.
			Those come from the following surveys:
				SV1 (blue, 37 tiles),
				One-Percent (red, 239 tiles),
				and Main (May-July 2021: green, 305 tiles; September-December 2021: orange, 973 tiles).
		}
		\label{fig:specobs}
	\end{center}
\end{figure}

We briefly summarize here the DESI ELG spectroscopic observations used hereafter; the interested reader can find details in \citet{desi-collaboration22b}.
DESI observations are conducted by 'tile`, i.e. by group of 5,000 fibered targets observed at once.
Each tile is observed so as to reach a required SNR value on average on all spectra.
This is done through the computation of an effective exposure time, $\efftime$, which accounts for observing conditions and of per-fiber properties \citep{guy22a}.
The sky map of the DESI ELG tiles used in this paper is displayed on Figure~\ref{fig:specobs}, which shows that each program has a specific tiling coverage of the footprint.

% figure: tiles

% sv1 obs
The above data contain 37 tiles with ELG targets from the SV1 (December 2020 to March 2021), which explored extended samples to finalize the target selection (see Section~\ref{sec:svts}).
Those typically have $\efftime \sim 4,000$s, i.e. four times the nominal Main survey \efftime, so that the observations provide secure data to study the faint end of the explored samples.
Three of those tiles have much higher $\efftime$ (7,000 -- 15,000s) and were used to build a truth table of about ten thousand ELG spectra with Visual Inspection (VI) \citep{lan22a}.

% sv3 obs
The One-Percent Survey (April 2021 to May 2021) observed 239 dark tiles distributed over 20 regions ('rosettes`) of the NGC with an \efftime~of about 1300s, i.e. 30 percent larger than the nominal Main survey \efftime.
A specificity of the One-Percent Survey observations is that most targets which did not have a conclusive $\zspec$ after a first observation were re-observed with another tile, to increase the SNR.
This significantly complicates the analysis in Section~\ref{sec:mainspec}, as repeat observations of the faintest targets to secure a reliable $\zspec$ measurement are not representative of the Main Survey.
In what follows, repeat observations -- i.e., observations of the same target from different tiles -- are thus removed from the One-Percent survey analysis.

% main obs
We use the Main Survey observations processed in \citet{guy22a}, which have been taken from May 2021 to July 2021, and include 305 dark tiles with a narrow distribution of \efftime~(1100s $\pm$ 190s).
This dataset, displayed in green in Figure~\ref{fig:specobs}, only covers part of the North and South-DECaLS footprints.
Lastly, for the redshift distribution (Figure~\ref{fig:nz}) and the comparison with the HSC $\zphot$ (Figure~\ref{fig:desi_hsc}), we complete this Main Survey sample with 973 Main dark tiles observed from September 2021 to December 2021 (in orange in Figure~\ref{fig:specobs}), which provide a significant coverage of the SGC, so that we have a representative sampling of the three footprints (North, South-DECaLS, and South-DES), in particular in terms of imaging depth, Galactic extinction, and stellar density,.
The pipeline reduction for that sample is not rigorously the same as that described in Section~\ref{sec:fujalupe} -- it is a slightly less advanced, but is a very close version.

% section: fuji
\subsection{Data reduction} \label{sec:fujalupe}
The spectroscopic data reduction and the redshift fitting with the \texttt{Redrock} software\footnote{\niceurl{https://github.com/desihub/redrock}} are fully described in \citet{guy22a} and \citet{bailey22a}, respectively.
We discard any observed spectrum with flagged issues in the data ($\texttt{COADD\_FIBERSTATUS}~!= 0$).
A key output quantifying the reliability of the best-fit $\zspec$ is the $\chi^2$ difference between the best-fit template and the second best one (\texttt{DELTACHI2}).
A large \texttt{DELTACHI2} generally implies a reliable $\zspec$ measurement.

We also use the SNR of the measured \oii flux (\texttt{FOII\_SNR}) computed as follows.
The continuum is estimated in the vicinity of the doublet, from the wavelengths 200 \AA~(in rest-frame) blue-wards of the \oii doublet.
Then the \oii doublet is simply fitted with two Gaussians at the expected positions corresponding to the measured $\zspec$.
The \oii flux, the line ratio, and the line width are let free in the fit.

\begin{figure*}[!h]
	\begin{center}
		\includegraphics[width=0.95\textwidth]{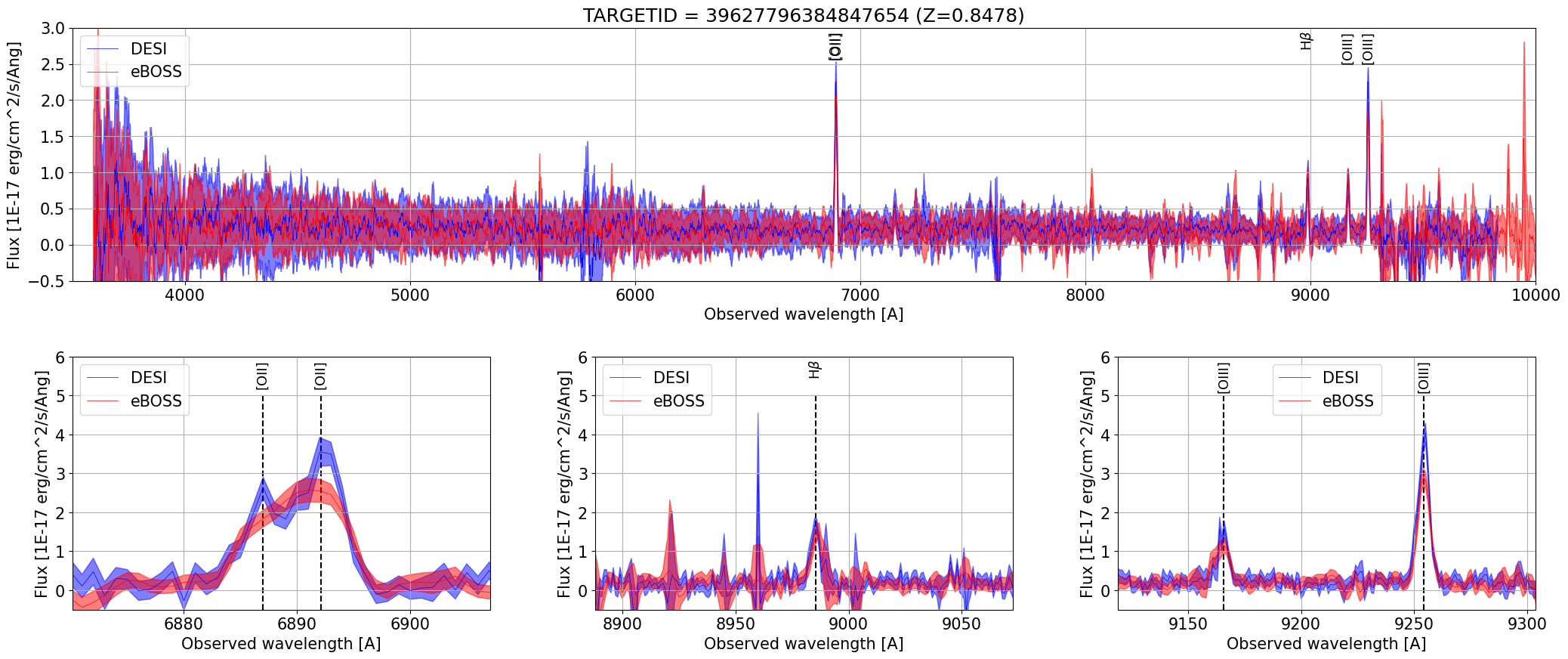}
		\caption{
			Typical ELG spectrum, observed with eBOSS in $\sim$1 hour (red) and DESI in $\sim$15 minutes (blue).
			The ELG target has $g_{\rm fib} = 23.2$ mag and $\zspec = 0.8478$.
			The top panel shows the full observed wavelength range, and the three bottom panels are zooms on the main emission lines for that spectrum: the \oii doublet (left), the H$\beta$ line (middle), and the \oiii $\lambda \lambda$ 4960,5008 \AA~lines (right).
			Shaded regions show the estimated 1$\sigma$ uncertainty on the measured flux.
			The spectra in the top panel are smoothed with a box of 11 pixels.
		}
		\label{fig:desi_sdss}
	\end{center}
\end{figure*}

Figure~\ref{fig:desi_sdss} compares the DESI spectrum of an ELG target to the one observed with the eBOSS survey.
The typical eBOSS observations were of one hour, against fifteen minutes for a typical DESI observation.
This is a representative $\zspec \sim 0.85$ ELG spectrum, with mostly undetected continuum and some significant emission lines.
The zoom panels on the bottom row show the improvement brought by DESI: its higher resolution allows one to nicely resolve the \oii doublet, which provides an unambiguous feature to estimate the redshift; besides it also provides sharper emission lines, with higher SNR. \AR{I could have picked one where the desi is much more better than the eboss, but I don t want to make eboss look bad!}

% figure: desi vs. sdss

% section: elg-qso weighting
\subsection{Weighting ELG and QSO targets}
Except for 12 ELG-dedicated SV1 tiles, the ELG targets always have a fiber assignment priority lower than the QSO and LRG targets.
The intersection between ELG and LRG target samples is virtually empty.
However, the ELG and QSO target samples intersection is significant: for instance, in the Main Survey, about one third of the QSO targets are also ELG targets, mostly ELG\_LOP ones; those $\sim$100 deg$^2$ targets represent about 5 percent of the ELG\_LOP targets.
Nevertheless, this 'ELGxQSO` subsample is not representative of the overall ELG sample, since it consists of bright objects with different colors.

As the QSO targets have top priority for the fiber assignment, the first passes will mostly be filled with QSO and LRG targets, and the ELG targets will be assigned during the later passes.
For the One-Percent Survey, as the program is finished, all ELG\_LOP targets were assigned.
But for the not-ELG-dedicated SV1 tiles or the -- currently not finished -- Main Survey, the 'ELGxQSO` targets are well over-represented in the observed spectra.
For instance, in a typical first pass tile of the Main survey, those 'ELGxQSO` targets can represent up to 30 percent of the observed ELG\_LOP targets.

In the analysis in Section~\ref{sec:mainspec} -- except for the repeat analysis, we thus correct that effect, and appropriately down-weight the 'ELGxQSO` observed targets.
We split the sky in healpix pixels large enough\footnote{\texttt{nside} = 16, i.e. a pixel area of 13.4 deg$^2$.} to reasonably track the selections density fluctuations.
For each of the SV1, One-Percent, and Main Survey, and each of the Main ELG\_LOP, and ELG\_VLO selections, we compute $f_{\rm targ}$ the fraction QSO targets in the considered selection for each pixel.
If we note $n_{\rm QSO}$ and $n_{\rm notQSO}$ the number of spectroscopically observed QSO and non-QSO targets for the considered selection in each pixel, we define the per-pixel weight to be applied to the $n_{\rm QSO}$ targets has follows:
$w_{\rm QSO} = f_{\rm targ} \times n_{\rm notQSO} / (n_{\rm QSO} - f_{\rm targ} \times n_{\rm QSO})$.
This ensures that $w_{\rm QSO} \times n_{\rm QSO} / (n_{\rm notQSO} + w_{\rm QSO} \times n_{\rm QSO}) = f_{\rm targ}$, i.e. that the weighted 'ELGxQSO` observed targets represent the same fraction of the observed sample than that of the parent target sample for each pixel.\\

%=======================================================
% Section : Spectroscopic properties of the Main Survey
%=======================================================
\section{Spectroscopic properties of the Main sample} \label{sec:mainspec}

This Section presents the spectroscopic properties of the ELG targets, based on the analysis of the spectroscopic data presented in the Section~\ref{sec:specdata}.

% section: reliable zspec criterion
\subsection{Reliable $\zspec$ criterion} \label{sec:zcrit}

% zcrit intro
We introduce the criterion used hereafter to select a reliable $\zspec$ for the ELG spectra, which is a cut in the \{\texttt{FOII\_SNR}, \texttt{DELTACHI2}\} space.
We emphasize that this is a preliminary, simple criterion, showing at first order what can be achieved.

% deltachi2-only not relevant
As for each DESI tracer, the ELG spectra require a dedicated reliable $\zspec$ measurement criterion, which maximizes the fraction of selected redshifts and minimizes the fraction of catastrophic redshifts in the selected sample, typically at the one percent level.
Such requirements are driven by the LSS analysis, which is sensitive to catastrophic redshifts \AR{reference?}.
One possible criterion is a high \texttt{DELTACHI2} value, as used for other DESI tracers \citep{hahn22a,zhou22a,chaussidon22a}.
The specificity of the ELG spectra is that they are at low SNR: a $\zspec$ reliably estimated with the \oii doublet may not have a large \texttt{DELTACHI2} value, as the pixels related to the \oii doublet represent a marginal fraction of the pixels and a fit with a single emission line with a different redshift could still provide a comparable $\chi^2$.
For that reason, selecting reliable $\zspec$ with a \texttt{DELTACHI2} criterion only would discard a large fraction of good redshifts, noticeably at high redshift, where the \oii doublet is the only feature in the spectrum.

\begin{figure}[!h]
	\begin{center}
		\begin{tabular}{c}
			\includegraphics[width=0.95\columnwidth]{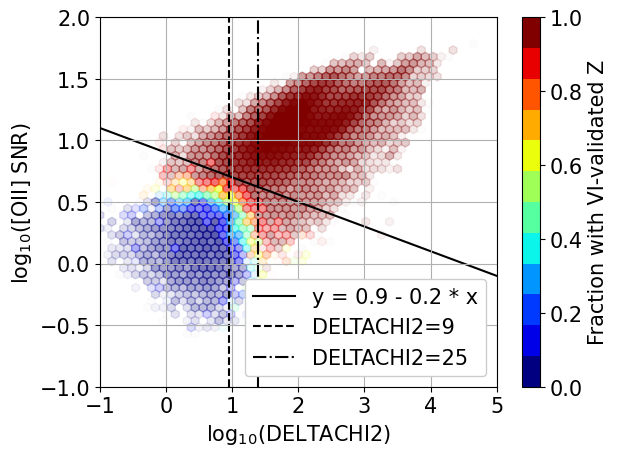}\\ 
			\\[1pt]
			\includegraphics[width=0.95\columnwidth]{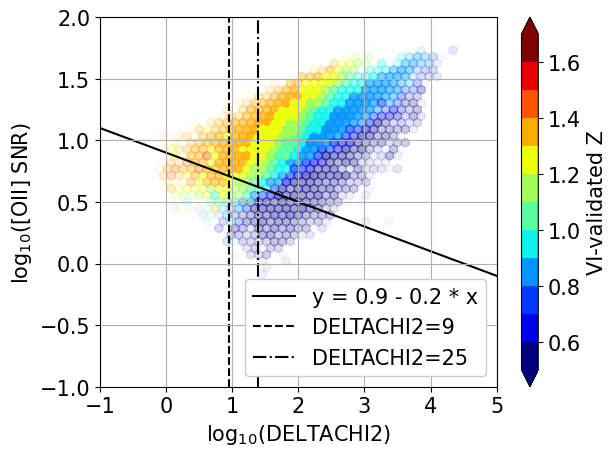}\\
		\end{tabular}
		\caption{
			ELG target properties in the \{\texttt{FOII\_SNR}, \texttt{DELTACHI2}\} plane.
			Top panel: fraction of $\zspec$ validated by VI.
			Bottom panel: average $\zspecvi$.
			On both plots, the slanted solid line is our criterion to select reliable $\zspec$ measurements, and the dashed, dot-dashed vertical lines illustrate two threshold values for a lower cut in \texttt{DELTACHI2}.
			The symbol transparency scales with the logarithm of the density.
			The data come from several reductions with $200\rm s < \efftime < 1600\rm s$ of the 3.5 thousand ELG\_LOP and ELG\_VLO targets with a VIed spectra (see text).
		}
		\label{fig:zcrit}
	\end{center}
\end{figure}

% zcrit
A relevant parameter space to consider in this regard is the \{\texttt{FOII\_SNR}, \texttt{DELTACHI2}\} space.
Figure \ref{fig:zcrit} shows the criterion we adopt in this paper:
% Equation: zcrit
\begin{equation}
	\rm{log}_{10} (\texttt{FOII\_SNR}) > 0.9 - 0.2 \times \rm{log}_{10} (\texttt{DELTACHI2})
	\label{eq:zcrit}
\end{equation}
This Figure is computed using approximately 3.5 thousand ELG\_LOP and ELG\_VLO VIed spectra.
For those spectra, the VI provides two informations from the deep reductions: $\zspecvi$ and $\qavi$, its confidence level; both those quantities are merged from the diagnosis of several inspectors.
The VI confidence level $\qavi$ ranges from 0 to 4 and, following the definition established in \citet{lan22a}, spectra with $\qavi \geq 2.5$ are considered to provide a robust $\zspecvi$, which we consider as the truth here.
We conservatively consider all spectra with $\qavi < 2.5$ to be a failure in shallower reductions.
As those observations come from the SV1 three deep, VIed tiles, which have been exposed with many exposures, we are able to generate several tens of coadded reductions with \efftime~ranging from 200s to 1600s.
Then, for each $\zspec$ measurement from those shallower reductions, we compare its value to the $\zspecvi$ from the deep reductions.
We consider that the $\zspec$ measurement is 'VI-validated` if it verifies the two following criteria:
% Equation: vi-validated
\begin{subequations}
	\begin{align}
		\qavi \geq 2.5 \label{eq:vizok_a}\\
		c \cdot |\zspec - \zspecvi| / (1 + \zspecvi) < 1000 \; \rm{km.s}^{-1}, \label{eq:vizok_b}
	\end{align}
\end{subequations}
where $c$ is the speed of light in km.s$^{-1}$.
The top panel of Figure~\ref{fig:zcrit} shows that a simple cut in \texttt{DELTACHI2} is not optimal at all:
 	a sample selected with the fiducial $\texttt{DELTACHI2} > 9$ threshold (used in the \texttt{Redrock} to flag a low-reliability $\zspec$ measurement), would be highly contaminated with catastrophic $\zspec$ measurements;
	a more conservative threshold, e.g., $\texttt{DELTACHI2} > 25$, would discard a significant number of reliable $\zspec$ measurements.
Our simple criterion of Equation~\ref{eq:zcrit} selects more than 95 percent of the reliable $\zspec$ measurements, while keeping a very low fraction of catastrophic $\zspec$ (about one percent).

% figure: zcrit

% z locii in the {foii_snr, deltachi2} space
The bottom panel of Figure~\ref{fig:zcrit} displays the average redshift for each position in the \{\texttt{FOII\_SNR}, \texttt{DELTACHI2}\} plane, using the $\zspecvi$ measurements with $\qavi \geq 2.5$.
It shows that the high-redshift ELG targets, which are the most valuable ones for DESI, have a low \texttt{DELTACHI2} value, despite their reliable $\zspec$ measurement (likely from the identification of the resolved \oii~doublet).

% foii_snr vs. z
Lastly, we illustrate in Figure~\ref{fig:foiiz} how $\zspec$ measurements selected by Equation~\ref{eq:zcrit} are distributed in the \{{\texttt{FOII\_SNR}, \texttt{Z}\} plane, for the $0.6 < z < 1.6$ range (top) and zooming in the $1.45 < z < 1.55$ range (bottom).
To have the largest sample size, this Figure displays about 600 thousands Main ELG spectra observed in the Survey Validation and Main Survey from coadded reductions with $800\rm s < \efftime < 1200\rm s$.
A noticeable feature is the drop in the measured \texttt{FOII\_SNR} for some $\zspec$ values, especially at $z > 1.5$.
Those are mostly due to sky line subtraction residuals.

To illustrate this point, sky line wavelengths converted into redshifts are displayed at the top of the plots, assuming the redshift to be that of an \oii doublet appearing at the same observed wavelength as the sky line i.e. $\zspec = \lambda_{\rm{sky}} / 3728 - 1$.
Future data reduction improvements (sky subtraction, \oii flux measurement) will be valuable for the analysis of the ELG sample, hopefully increasing the fraction of reliable $\zspec$ at $z > 1.5$.

% figure: foii_snr vs. z
\begin{figure}[!h]
	\begin{center}
		\begin{tabular}{c}
			\includegraphics[width=0.95\columnwidth]{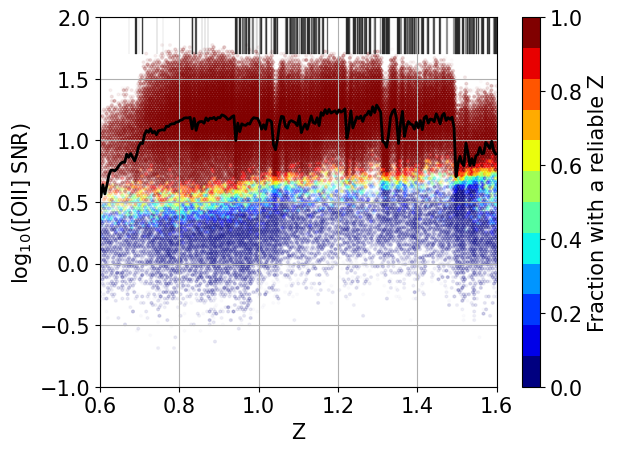}\\ 
			\\[1pt]
			\includegraphics[width=0.95\columnwidth]{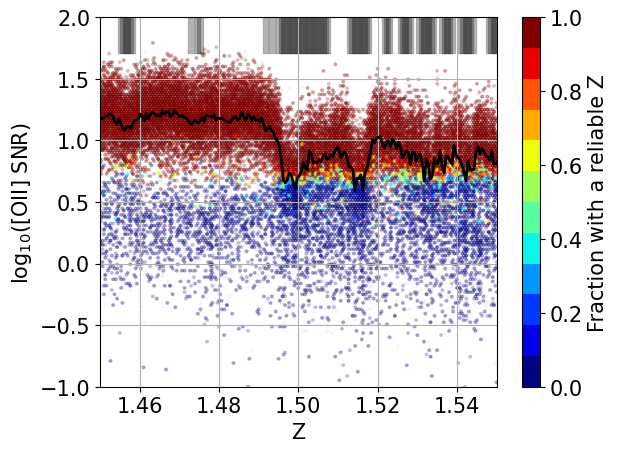}\\
		\end{tabular}
		\caption{
			Distribution of the fraction of reliable $\zspec$ measurements in the \{\texttt{FOII\_SNR}, \texttt{Z}\} plane, for the $0.6 < z < 1.6$ range (top) and zooming in the $1.45 < z < 1.55$ range (bottom).
			The solid black line displays the median \texttt{FOII\_SNR} value.
			The symbol transparency scales with the logarithm of the density.
			On both plots, the sky emission lines are displayed as black rectangles at the top of the plots, where the sky wavelength $\lambda_{\rm{sky}}$ is converted to $\zspec = \lambda_{\rm{sky}} / 3728 - 1$, that is to the redshift of an \oii doublet observed at that wavelength.
			The data come from about 600 thousand Main ELG spectra observed in coadded reductions with $800\rm s < \efftime < 1200\rm s$.
		}
		\label{fig:foiiz}
	\end{center}
\end{figure}

% future possible improvements
A detailed analysis of all the recent Main Survey spectra -- possibly enhanced with additional VI -- will allow to refine this Equation~\ref{eq:zcrit} criterion.
Likely improvements would be to refine the cut in the space to enlarge the fraction of selected $\zspec$; to refine the selection in the $1.5 < z < 1.6$ range, where the \oii doublet falls in the forest of sky lines; or refine it at $z < 1$, using other lines, like the \oiii $\lambda \lambda$ 4960,5008 \AA.

% section: catastrophics
\subsection{Fraction of selected catastrophic $\zspec$ estimated from VI}\ \label{sec:catavi}
An important quantity to control is the fraction of $\zspec$ selected with our reliability criterion (Equation~\ref{eq:zcrit}) which has a catastrophic $\zspec$ estimate.

% vi sample for cata 
We first assess this fraction with using the SV1 VI sample from the three deep ELG tiles.
We identify as a catastrophic $\zspec$ estimate any measurement passing our Equation~\ref{eq:zcrit} criterion, but failing Equation~\ref{eq:vizok_a} or Equation~\ref{eq:vizok_b}.
We restrict here to spectra that would be selected for an LSS analysis, i.e. redshifts in the $0.6 < z < 1.6$ range passing our criterion (about 2.9 thousand ELG\_LOP targets and 0.7 thousand ELG\_VLO targets).
From the multiple reductions, we have at hands about sixty reductions with spanning \efftime~values between 200s and 1600s.

% cata: figure
Figure~\ref{fig:cata_vi} presents for the ELG\_LOP sample the fraction of catastrophic $\zspec$ as a function of \efftime~values (top) and  $\zspec$ (bottom), and demonstrates that our criterion is effective in keeping the catastrophic $\zspec$ fraction at the order of the percent level.
For a typical $\efftime \sim 1000\rm s$ -- the nominal $\efftime$ for the Main Survey, the catastrophic $\zspec$ fraction for the ELG\_LOP sample is $\sim$0.2 percent; for the ELG\_VLO sample, it is virtually zero.
The fraction is independent of \efftime; this is the desired behavior, i.e. a shallow reduction would naturally select much less spectra, but those would be of similar quality as spectra from a deeper reduction.
As a function of $\zspec$, the catastrophic $\zspec$ fraction is very low for $z < 1.2$, but starts to increase for $1.3 < z < 1.4$, and is more significant for $1.5 < z < 1.6$.

The reasons are twofold.
First, this reflects the fact that as redshift increases, emission lines move redward leaving only the \oii doublet.
Then, for $1.5 < z < 1.6$, the \oii~doublet falls in a region with many strong sky lines subtraction residuals (see Figure~\ref{fig:foiiz}), likely preventing the visual inspectors to securely confirm the redshift; our conservative choice to consider all redshifts from a target with $\qavi < 2.5$ as a failure (see Equation \ref{eq:vizok_a}) likely explains the high values of catastrophic $\zspec$ fraction in $1.5 < z < 1.6$ (see next Section).
In any case, to reduce the catastrophics $\zspec$ fraction in the $1.5 < z < 1.6$ range will require to improve the sky subtraction in the reduction pipeline.\\

% figure: catastrophics vi
\begin{figure}[!h]
	\begin{center}
		\begin{tabular}{c}
			\includegraphics[width=0.95\columnwidth]{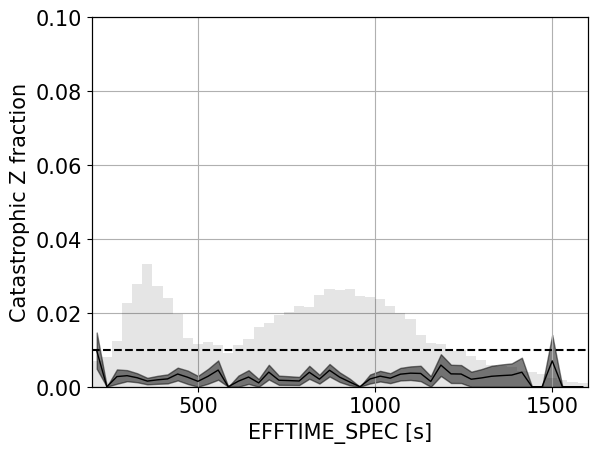}\\ 
			\\[1pt]
			\includegraphics[width=0.95\columnwidth]{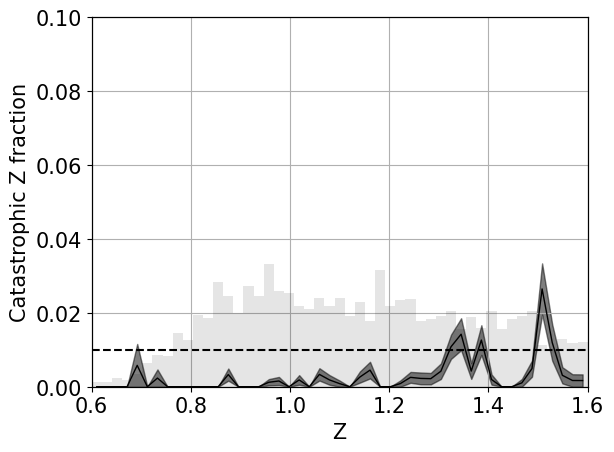}\\
		\end{tabular}
		\caption{
			Fraction of catastrophic $\zspec$ measurements for Main ELG\_LOP targets with $0.6 < z < 1.6$ and a reliable $\zspec$ (Equation~\ref{eq:zcrit}), as a function of $\efftime$ (top) and $\zspec$ (bottom).
			The grey histograms show the distribution of the probed values and the dashed line highlights a catastrophic $\zspec$ fraction of one percent.
			The data come from different reductions of the 2.9 thousand ELG\_LOP targets with VI.
		}
		\label{fig:cata_vi}
	\end{center}
\end{figure}

% repeat sample for cata
\subsection{Fraction of selected catastrophic $\zspec$ and $\zspec$ precision estimated from repeats}\ \label{sec:catarp}
We use repeat observations of the Main ELG targets to re-assess the catastrophic $\zspec$ fraction with an independent method, and to determine the $\zspec$ measurement precision.

We build a sample of about 19 (5) thousand pairs of independent repeat observations of the Main ELG\_LOP (ELG\_VLO) targets as follows.
We consider the SV1, One-Percent, and Main per-night reductions with $800\rm s < \efftime < 1200\rm s$, each reduction being made from independent observations (either different tile observation or different exposures for a given tile).
We then restrict to the Main ELG targets with a reliable $\zspec$ measurement (Equation~\ref{eq:zcrit}) and with one of the two measurements in $0.6 < z < 1.6$, and identify targets having two or more reductions.
For each pair, we consider the redshift difference $dv = c \cdot (z_0 - z_1) / (1 + z_0)$, where $c$ is the speed of light in km.s$^{-1}$.

First, the Figure~\ref{fig:cata_rp_z} independently re-assesses with these repeat observations the catastrophic $\zspec$ fraction as a function of redshift.
We obtain sub-percent fractions for all redshifts.
This is consistent with the Figure~\ref{fig:cata_vi}, except for the $1.3 < z < 1.6$ range, where the fraction from the repeats has much less pronounced peaks.
That is consistent with our statement in previous section, that those peaks in Figure~\ref{fig:cata_vi} are likely driven by our conservative choice to consider all redshifts from a target with $\qavi < 2.5$ as a failure.

% figure: catastrophics repeats = f(redshift)
\begin{figure}[!h]
	\includegraphics[width=0.95\columnwidth]{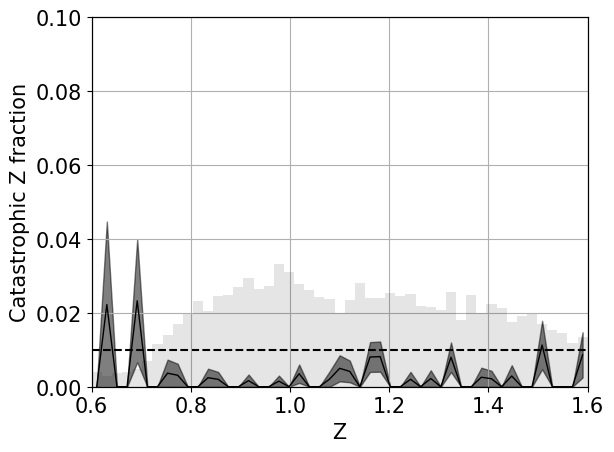}\\ 
	\caption{
		Same as Figure~\ref{fig:cata_vi}, but with data coming from independent repeat observations from the SV1, One-Percent and Main surveys (see text for more details).
		}
	\label{fig:cata_rp_z}
\end{figure}

The top panel of Figure~\ref{fig:repeat} shows $dv$ as a function of \texttt{FOII\_SNR} for the ELG\_LOP sample: only 0.2 percent of the pairs have a catastrophic measurement (red dots); the ELG\_VLO sample has virtually zero catastrophic measurements.
This fraction is in agreement with the catastrophic rate estimated in Section~\ref{sec:catavi} from a totally independent method.

Besides, this panel emphasizes that the \oii doublet is crucial for the ELG $\zspec$ measurement, as it shows a clear correlation between the redshift precision and the \texttt{FOII\_SNR}.
The $dv$ distribution is reported in the bottom panel of Figure~\ref{fig:repeat}, and can be reasonably modeled by the weighted sum of two Gaussian distributions centered on zero with widths of 5 and 20 km.s$^{-1}$.
Eventually, if we use the same statistical measurement as \citet{lan22a} we measure for the ELG\_LOP (ELG\_VLO) sample $\rm{MAD}(dv) \times 1.48 / \sqrt{2} \sim 7 \; \rm{km.s}^{-1}$ ($9 \; \rm{km.s}^{-1}$), where MAD is the median absolute deviation, in agreement with \citet{lan22a}.
This $\zspec$ precision is the best among DESI tracers \citep{lan22a,alexander22a}, as the $\zspec$ is based on sharp emission lines.

% figure: repeat (cata, precision)
\begin{figure}[!h]
	\begin{center}
		\begin{tabular}{c}
			\includegraphics[width=0.95\columnwidth]{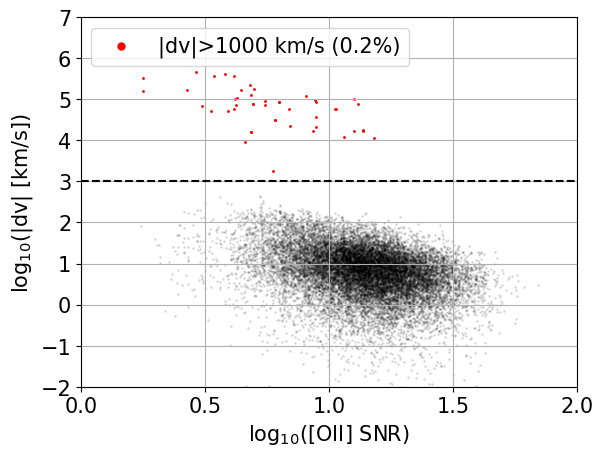}\\ 
			\\[1pt]
			\includegraphics[width=0.95\columnwidth]{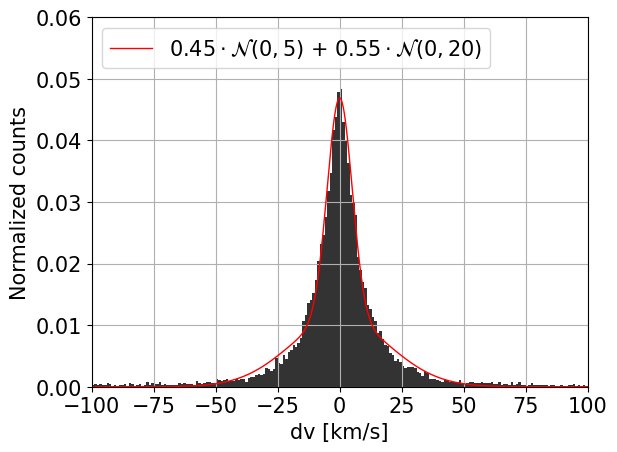}\\
		\end{tabular}
		\caption{
			Redshift difference $dv$ for 19 thousand pairs of independent repeat observations of Main ELG\_LOP targets with $0.6 < z < 1.6$.
			\textit{Top panel}: $dv$ as a function of \texttt{FOII\_SNR}; the black dashed line ($dv = 1000 \; \rm{km.s}^{-1}$) illustrates the threshold used to identify the pairs with catastrophic measurement (red dots).
			\textit{Bottom panel}: normalized distribution of the $dv$ values; the red curve illustrates a reasonable representation of the distribution, made of the weighted sum of two Gaussian distributions with widths of 5 km.s$^{-1}$ and 20 km.s$^{-1}$.
			The data come from independent repeat observations from the SV1, One-Percent and Main surveys (see text for more details).
		}
		\label{fig:repeat}
	\end{center}
\end{figure}

% section: efficiency = f(efftime_spec)
\subsection{Efficiency: fraction of selected $\rm{z_{min}} < \zspec < \rm{z_{max}}$} \label{sec:selfrac} 
We present the fraction of ELG spectra which is selected by our reliability criterion from Equation~\ref{eq:zcrit}.
Figure~\ref{fig:selfrac} displays that fraction as a function of \efftime, for the Main ELG\_LOP and ELG\_VLO samples, and for the three surveys (SV1, One-Percent, and Main; top panel).
Combining those three surveys (bottom panel) allows us to probe a wide range of \efftime, SV1 exploring the entire range but with low statistics, One-Percent probing values from 1000s to 1500s, right in the high tail of the Main survey range.
For this Figure, we restrict to reductions with $100\rm s < \efftime < 1600\rm s$.
As demonstrated in Section~\ref{sec:photsyst}, the ELG\_LOP and ELG\_VLO selections have some dependencies with the $g$-band imaging depth, the Galactic extinction, and the distance to the Sagittarius Stream.
To allow a comparison of the three surveys, we:
	(1) restrict to regions with $g$-band depth smaller than 24.5 mag and E(B-V) smaller than 0.1 mag;
	(2) subsample the SV1 and One-Percent data so that the distance to the Sagittarius Stream values are representative of the distribution probed by the Main Survey.

% figure: efficiency = f(efftime_spec)
\begin{figure}[!h]
	\begin{center}
		\begin{tabular}{c}
			\includegraphics[width=0.95\columnwidth]{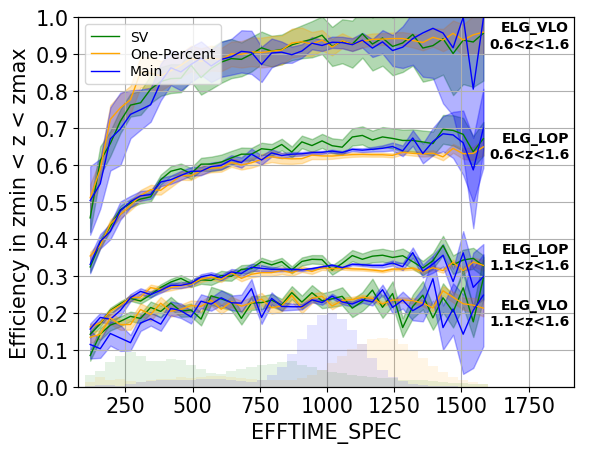}\\ 
			\\[1pt]
			\includegraphics[width=0.95\columnwidth]{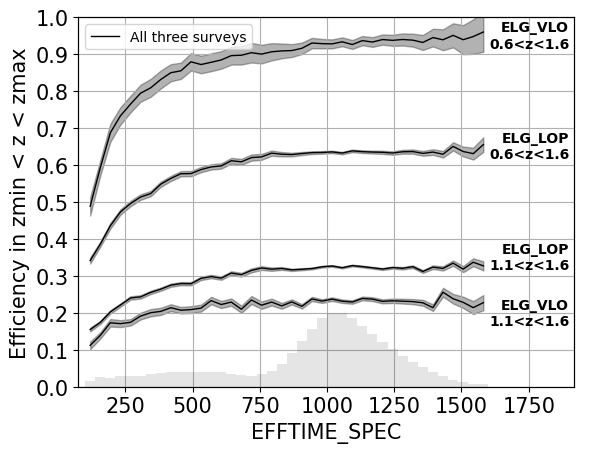}\\
		\end{tabular}
		\caption{
			ELG Main selection efficiency in the $0.6 < z < 1.6$ and $1.1 < z < 1.6$ ranges, as function of EFFTIME\_SPEC, for the three surveys (Survey Validation SV1 and One-Percent Survey, and Main Survey).
			\textit{Top panel}: split per survey; \textit{Bottom panel}: all surveys together.
			The histograms show the $\efftime$ normalized distributions.
		}
		\label{fig:selfrac}
	\end{center}
\end{figure}

% selfrac: efficiency
We call efficiency for a given $z_{\rm{min}} < z <z_{\rm{max}}$ range the fraction of observed ELG targets which obtain a reliable $\zspec$ measurement in that redshift range.
We display in this figure the efficiency for the two important redshift ranges to control for the target sample, namely $0.6 < z < 1.6$, the nominal redshift range, and $1.1 < z < 1.6$, the high-redshift part of that range, where ELGs are the most important tracers for DESI.
For the Main Survey nominal $\efftime \sim 1000$s, the ELG\_LOP selection has an efficiency in $0.6 < z < 1.6$ of 60-65 percent, and an efficiency in $1.1 < z < 1.6$ of 30-35 percent.
The ELG\_VLO selection has a much higher efficiency in $0.6 < z < 1.6$ of 90-95 percent, but an efficiency of only 20-25 percent in $1.1 < z < 1.6$, as expected from the designed photometric cuts.

% selfrac: consistency among surveys
A noticeable feature in Figure~\ref{fig:selfrac} is the overall agreement between the three surveys.
This highlights the relevance of DESI approach, with using the SV1 to explore a large selection sample, then One-Percent to refine the selections and check them with observations slightly deeper than nominal, before finalizing the Main survey selection.
And this is what allows us to combine the three surveys together (bottom panel of Figure~\ref{fig:selfrac}), leading to a precise measurement over a large range of $\efftime$ values.

% selfrac: variation with efftime
The second visible feature is the flattening of the efficiency curves towards large \efftime~values.
As expected, the efficiency is low for low \efftime~values, as the ELG spectra do not have enough SNR to securely measure a $\zspec$ value for most of the sample.
Then the efficiency strongly increases with increasing \efftime~up to approximately 750s.
Finally the efficiency flattens for \efftime~values larger than 750s.
As a consequence, the efficiency is rather constant over the \efftime~range probed by the Main survey (blue histogram in the top panel), which naturally has some scatter around the requested value of 1000s.

% section: efficiency = f(mag, color)
\subsection{Efficiency in the photometric space} \label{sec:effphot}
This Section analyzes the redshift efficiency in the $0.6 < z < 1.6$ and $1.1< z < 1.6$ ranges as a function of the $g$-band magnitude and of the $grz$-band colors.
We use the SV1 ELG sample in order to quantify how the efficiency varies in the photometric space for the Main ELG selection, and at the borders of that selection.
Such a study was used to finalize the Main Survey ELG selection.

We use reductions of the 25 SV1 ELG-only tiles with $800\rm{s} < \efftime < 1200\rm{s}$, i.e. \efftime~values representative of the Main Survey; those tiles include about 43.2 thousand ELG targets.
The results are presented in Figure~\ref{fig:effphot}, where the top (resp. bottom) row is the efficiency in the $0.6 < z < 1.6$ (resp. $1.1 < z < 1.6$) range.
The left column shows the $g-r$ vs. $r-z$ diagram, the middle column shows the $\coii$ vs. $g_{\rm{fib}}$ diagram, and the right column shows the $g_{\rm{tot}}$ vs. $g_{\rm{fib}}$ plane.
On all plots, the Main ELG\_LOP selection is displayed as a solid dark line, the Main ELG\_VLO selection as a dashed red line, and the SV1 ELG selection as a thick solid grey line.

% figure: efficiency = f(mag, color)
\begin{figure*}[!h]
	\begin{center}
		\begin{tabular}{ccc}
			\includegraphics[width=0.3\textwidth]{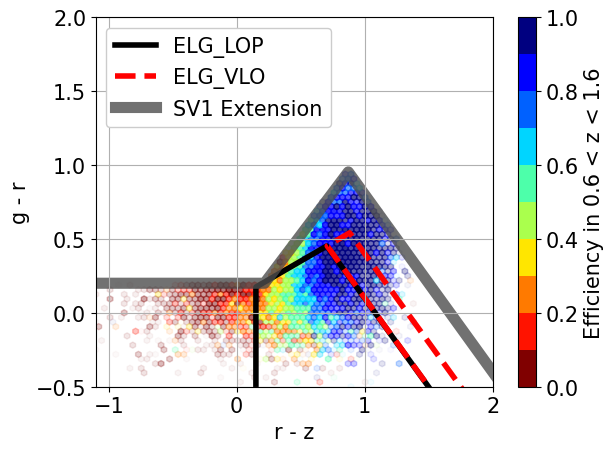} &
			\includegraphics[width=0.3\textwidth]{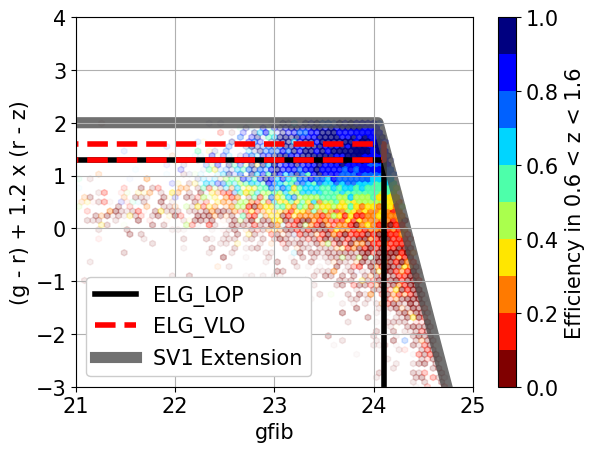} &
			\includegraphics[width=0.3\textwidth]{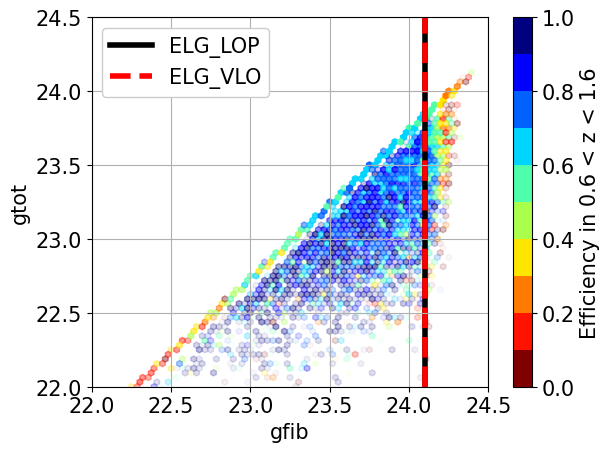}\\
			\\[1pt]
			\includegraphics[width=0.3\textwidth]{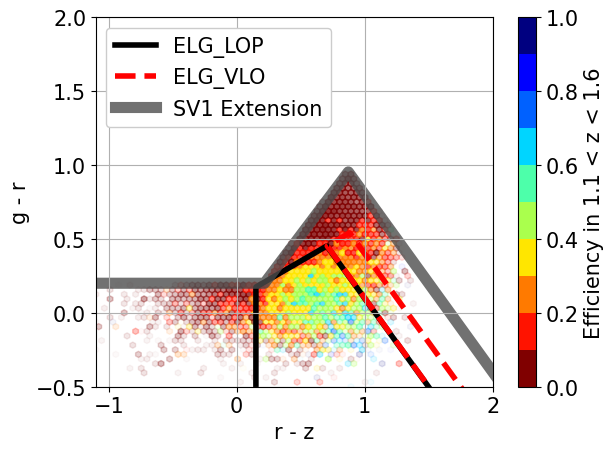} &
			\includegraphics[width=0.3\textwidth]{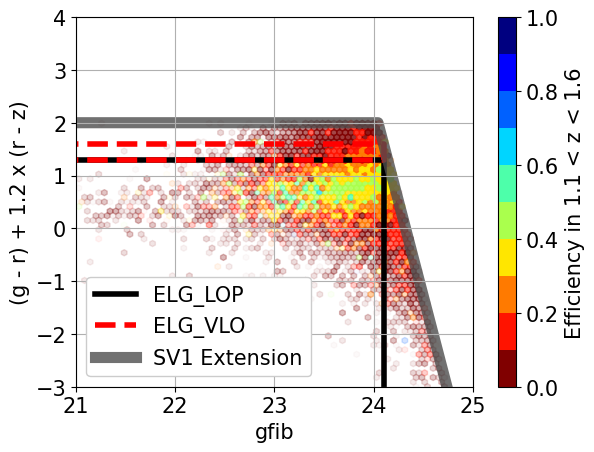} &
			\includegraphics[width=0.3\textwidth]{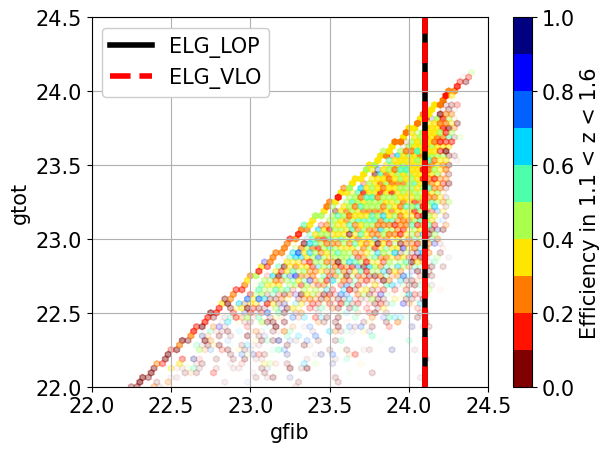}\\
		\end{tabular}
		\caption{
			Efficiency as a function of colors and magnitude, in the $0.6 < z < 1.6$ (top) and $1.1 < z < 1.6$ (bottom) range.
			The left column shows the $g-r$ vs. $r-z$ diagram, the middle column shows the $\coii$ vs. $g_{\rm{fib}}$ diagram, and the right column shows the $g_{\rm{tot}}$ vs. $g_{\rm{fib}}$ plane.
			The Main ELG\_LOP and the ELG\_VLO selections are displayed as solid black lines and dashed red lines, respectively; the SV1 ELG selection is displayed as a thick solid gray line.
			The data come from different reductions of SV1 ELG-only tiles, with $800\rm s < \efftime < 1200\rm s$, representative of the Main Survey $\efftime$ values.
		}
		\label{fig:effphot}
	\end{center}
\end{figure*}

% effphot: grz
The $g-r$ vs. $r-z$ plots include a $g_{\rm{fib}} < 24.1$ selection, i.e. the same magnitude limit as the Main ELG sample.
The efficiency in the $1.1 < z < 1.6$ interval (bottom) was the key motivation behind the Main ELG selection box definition.
The efficiency inside the ELG\_LOP selection box is rather stable, but sharply drops on the blue $r-z$ side (likely because any $z > 1.6$ galaxy cannot provide a reliable $\zspec$) and on the red $g-r$ side (likely due to contamination from stars and $z<0.6$ galaxies).
On the red $r-z$ side, the efficiency drops more smoothly, which justifies the ELG\_VLO selection box definition, in conjunction with the very high efficiency of that $g-r$ vs. $r-z$ region for the $0.6 < z < 1.6$ range (top plot).

% effphot: coii
The $\coii$ vs. $g_{\rm{fib}}$ diagram allows us to test the effect of selecting ELG targets fainter than $g_{\rm{fib}} < 24.1$.
At fixed $\coii$, the efficiency is actually rather stable for $\coii > 0$ when going to magnitudes fainter than the Main ELG cut (solid black line), illustrating the good performance of DESI.
Nevertheless, such a selection would also bring in many failures from the $\coii < 0.5$ region, which would mitigate the gain in efficiency; that, combined with the fact that going fainter increases the density variations with imaging and foreground properties (see Section~\ref{sec:photsyst}), explains why it was not considered in the end.

% effphot: gfib vs. gtot
The  $g_{\rm{tot}}$ vs. $g_{\rm{fib}}$ plot include the $g-r$ vs. $r-z$ selection of the Main ELG cuts.
They confirm the expectations that a total magnitude based selection would add poor efficiency targets, likely extended targets with little flux inside DESI fibers.
This justifies our choice to use a fiber magnitude-based cut for the Main selection.

% section: n(z)
\subsection{Redshift distribution} \label{sec:nz}
We present the Main Survey ELG sample redshift distribution.
We consider the Main Survey observations up to December 2021, as this allows us to build a sample observed over a footprint fairly representative of the full DESI footprint (see Section \ref{sec:specobs}).
We restrict to tiles with $800\rm s < \efftime < 1200\rm s$.
This sample contains about 1.5 million ELG\_LOP spectra (North: 0.2 million, South-DECaLS: 1 million, South-DES: 0.3 million) and about 187 thousand ELG\_VLO spectra (North: 21 thousand; South-DECaLS: 132 thousand; South-DES: 34 thousand).

% nz: efficiency
For each sample and each footprint, Table~\ref{tab:eff} lists  the overall efficiency, i.e. the fraction of observed ELG spectra which provide a reliable $\zspec$ (Equation~\ref{eq:zcrit}), and the efficiencies in the $0<z<0.6$, $0.6<z<1.1$, and $1.1<z<1.6$ ranges.
With our current criterion, 68 to 73 percent of the ELG\_LOP targets provide a reliable $\zspec$ ($\sim$70 percent in North and South-DECaLS, 73 percent in South-DES).

% table: efficiencies
\begin{table*}
	\centering
	\begin{tabular}{lcccccc}
		\hline
		\hline
		Sample & Footprint & Target density & \multicolumn{4}{c}{Efficiency}\\
		 &  & [deg$^{-2}$] & All redshifts & $0 < z < 0.6$ & $0.6 < z < 1.1$ & $1.1 < z < 1.6$\\
		\hline
		\multirow{3}{*}{ELG\_LOP} & North & 1930 &0.71 & 0.05 & 0.33 & 0.32\\
		& South-DECaLS & 1950 & 0.68 & 0.05 & 0.29 & 0.34\\
		& South-DES & 1900 & 0.73 & 0.02 & 0.31 & 0.39\\
		\hline
		\multirow{3}{*}{ELG\_VLO} & North & 410 & 0.93 & 0.01 & 0.70 & 0.21\\
		& South-DECaLS & 490 & 0.94 & 0.01 & 0.67 & 0.25\\
		& South-DES & 480 & 0.95 & 0.00 & 0.69 & 0.26\\
	        \hline
    \end{tabular}
	\caption{
		Efficiency per sample and footprint for the Main Survey ELG selection.
		We call efficiency for a given $\rm{z_{min}} < z < \rm{z_{max}}$ range the fraction of observed ELG targets which provides a reliable $\zspec$ measurement (according to Equation~\ref{eq:zcrit}) in that redshift range.
	}
	\label{tab:eff}
\end{table*}

% nz: description
Figure~\ref{fig:nz} shows the expected density of observed ELG targets providing a reliable $\zspec$ at the end of the survey, when all passes will have been completed.
The distributions are normalized to
	the target density, multiplied by
	the fiber assignment rate at the end of the survey, multiplied by
	the fraction of observed ELG targets providing a reliable $\zspec$.
The target densities are reported in Table~\ref{tab:imagdens}.
The fiber assignment rate at the end of survey is expected to be 0.69 for the ELG\_LOP selection and 0.42 for the ELG\_VLO one \citep{raichoor22b,desi-collaboration22b}; those values account for the 1\% loss rate affecting the ELG observations, i.e. where observations are discarded because of non-valid fibers (e.g., due to mechanical issue, petal-rejection; see \citealt{desi-collaboration22b,schlafly22a}).
The fractions of observed ELG targets providing a reliable $\zspec$ are listed in the fourth column of Table~\ref{tab:eff}.
Altogether, the normalization for the three footprints range from 910 to 960 deg$^{-2}$ for the ELG\_LOP selection, and from 160 to 190 deg$^{-2}$ range for the ELG\_VLO selection.

% figure: n(z)
\begin{figure}[!h]
	\begin{center}
		\begin{tabular}{c}
			\includegraphics[width=0.95\columnwidth]{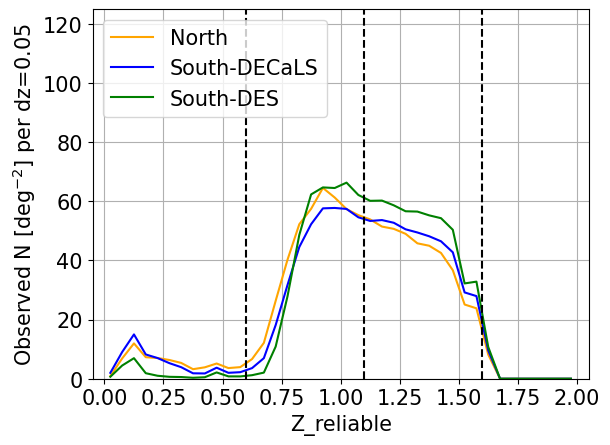}\\
			\\[1pt]
			\includegraphics[width=0.95\columnwidth]{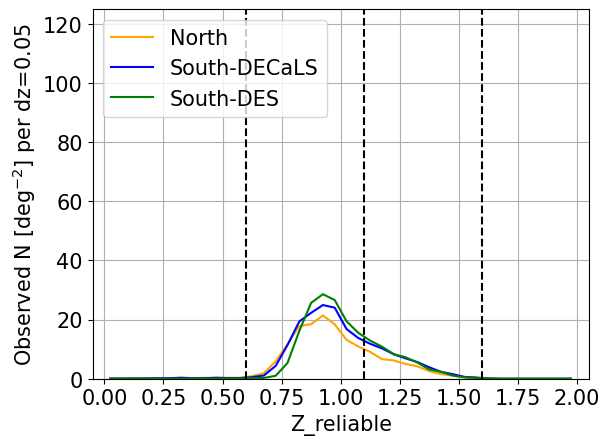}\\
		\end{tabular}
		\caption{
			Redshift distributions for the Main ELG\_LOP sample (top) and ELG\_VLO sample (bottom), split by footprint.
		The reported densities are the expected densities of observed ELG targets providing a reliable $\zspec$ (see text for more details).
		The vertical dashed lines emphasize the relevant redshift ranges for the ELGs: $z=0.6$, $z=1.1$, and $z=1.6$.
		}
		\label{fig:nz}
	\end{center}
\end{figure}

% nz: elg_lop discussion
Considering the efficiencies per redshift range in Table~\ref{tab:eff}, the ELG\_LOP redshift distribution provides more than 400 deg$^{-2}$ reliable $\zspec$ in both the $0.6<z<1.1$ and the $1.1<z<1.6$ ranges, and thus fulfills DESI requirements.
As expected, the deeper the imaging is, the more reliable $\zspec$ are gathered in the $1.1<z<1.6$ range.
The North and South-DECaLS footprints show comparable $z<0.6$ contamination of 5 percent, but the South-DECaLS footprint provides 30 deg$^{-2}$ more $z>1.1$ redshifts.
The South-DES footprint, with imaging 0.5 mag or more deeper than in DECaLS-South, has significantly better performances, with almost no $z<0.6$ contamination, and should bring 20 deg$^{-2}$ more $0.6<z<1.1$ and 60 deg$^{-2}$ more $1.1<z<1.6$ reliable redshifts than the South-DECaLS footprint, despite an overall target density smaller by 50 deg$^{-2}$.

% nz: elg_lop failures
Lastly, we stress that characterizing the redshift distribution of the $\sim$30 percent targets which do not provide a reliable $\zspec$ will be an important topic for the DESI ELG LSS analysis; in particular, estimating the fraction and distribution in the redshift range of interest.
This can be achieved with using for instance accurate $\zphot$ or the clustering redshift method \citep[e.g.][]{newman08a}.

We note that this fraction could be decreased thanks to further developments.
First, it is very likely that some failures are spurious targets close to bright or medium stars, as the angular masking was purposely chosen to be very minimal at the targeting step.
Preparatory work has shown that the target selection has indeed some overdensity close to bright or medium stars.
Another potential improvement could come from the pipeline, with for instance a better sky subtraction.
Lastly, as already said, the reliable $\zspec$ criterion could be refined.
Nevertheless, we expect that fraction of redshift failures to remain non-negligible in the end.
The VI analysis done on deep exposures showed that about 14 percent of the ELG\_LOP selection do not provide a VI-reliable $\zspec$; even if the VI was done on data processed with a less advanced reduction pipeline, it gives the order of magnitude of the effect.

% nz: elg_vlo discussion
The bottom panel of Figure~\ref{fig:nz} displays the redshift distribution of the ELG\_VLO sample.
As seen in Section~\ref{sec:photsyst}, this selection is less sensitive to the imaging depth, leading to less significant differences in the redshift distribution among the three footprints.
Nevertheless, their performances are still in the same order, the South-DES footprint providing the cleaner, higher-redshift sample, and the North footprint performing less well at high-redshift.
For all footprints, the ELG\_VLO selection should provide about 130 deg$^{-2}$ reliable redshifts in the $0.6<z<1.1$ range and about 50 deg$^{-2}$ reliable redshifts in the $1.1<z<1.6$ range, mostly because of the low fiber assignment rate of that sample (0.42).\\

%=======================================================
% Section : Conclusion
%=======================================================
\section{Conclusion} \label{sec:concl}

The ELG sample will constitute one-third of the 40 million extra-galactic DESI redshifts and will be used to probe the Universe over the $0.6 < z < 1.6$ range, and in particular over the $1.1 < z < 1.6$ range, where it will bring the tightest of the DESI cosmological constraints.
We presented the DESI ELG target selection used for survey validation and the final selection that was derived from it for the Main Survey.
% overview
The Main Survey ELG selection is composed of two disjoint sets of cuts, the ELG\_LOP and the ELG\_VLO selections, which have target densities of about 1940 deg$^{-2}$ and 460 deg$^{-2}$, respectively.
The ELG\_LOP sample, which has a higher fiber assignment priority, favors the $1.1 < z < 1.6$ range, whereas the ELG\_VLO sample, at lower fiber assignment priority, favors the $0.6 < z < 1.1$ range.
These two samples are completed by the ELG\_HIP sample, a 10 percent random subsample of the ELG\_LOP and ELG\_VLO samples, which has the same fiber assignment priority as the LRG one.

% ts
The target selection is based on the $grz$-band photometry rom the DR9 release of the Legacy Surveys.
As the ELG targets are at low SNR in the imaging, we define three footprints, isolating the three regions linked to the underlying observing programs with different imaging depths: North, South-DECaLS, and South-DES.
Both Main Survey ELG\_LOP and ELG\_VLO samples are selected with a $g$-band fiber magnitude cut $g_{\rm{fib}} < 24.1$, which favors \oii emitters and higher successful $\zspec$ measurement rates, and a specific $(g-r)$ vs.\ $(r-z)$ color box, which primarly selects the redshift range.
The Survey Validation SV1 ELG sample, which was used to tune the Main Survey cuts is an extended version of those, noticeably towards:
	(1) bluer $r-z$ targets, exploring the $1.1 < z < 1.6$ range;
	(2) redder $r-z$ targets, exploring the $0.6 < z < 1.1$ range;
	(3) fainter blue targets
	(4) $g$-band total magnitude selected targets.

% phot properties
We then presented the photometric properties of the Main Survey ELG selection.
In terms of magnitude, the ELG\_LOP sample is 0.2 mag fainter in the $r$-band and 0.5 mag fainter in the $z$-band than the ELG\_VLO sample, because of the different selection boxes in the $g-r$ vs. $r-z$ diagram.
For both samples, the target density is slightly different in each footprint, mostly because of the difference in imaging depth: ELG\_LOP ranges from 1900 to 1950 deg$^{-2}$ and ELG\_VLO from 410 to 490 deg$^{-2}$.
The imaging and foreground maps causing the largest target density fluctuations are the imaging depth, in particular in the $g$-band, and the Galactic dust extinction.
Attempts to correct those fluctuations are deferred to subsequent papers, when a cleaner sample will have been defined, e.g., after inclusion of appropriate angular masking and correction for $\zspec$ measurement failures.

% spec properties
Lastly, we presented the spectroscopic properties of the Main Survey ELG selection.
% data
For that purpose, we used observations from three DESI surveys covering two phases of validation and the first 7 months of the Main Survey.
We define a preliminary criterion to select reliable $\zspec$ measurements, that requires a minimal \oii doublet flux SNR as a function of the $\chi^2$ difference between the first and second redshift values that best fit the observed spectrum.
This criterion exploits the fact that the \oii doublet is the key emission line to measure accurate redshifts of star-forming ELGs.
% zcrit
Using visually inspected (VI) spectra tiles and repeat observations, we demonstrate that such a criterion is extremely efficient, since it selects most of the VI-confirmed $\zspec$ and keeps the fraction of catastrophic $\zspec$ measurements below one percent.
Nevertheless, this discards about 30 percent of the observed ELG\_LOP spectra (and 6 percent of the ELG\_VLO ones): even if some improvement in the data reduction and in the reliability criterion could reduce those percentages, we expect that it will remain non-negligible for the ELG\_LOP sample, and it will be necessary to characterize the redshift properties of the discarded spectra.

% elg_lop efficiency
We defined the efficiency in a given redshift range as the fraction of observed ELG spectra providing a reliable $\zspec$ in that redshift range.
Depending on the footprint, the ELG\_LOP selection has an efficiency of 2 to 5, 29 to 33, and 32 to 39 percent in the $0 < z < 0.6$, $0.6 < z < 1.1$, and $1.1 < z < 1.6$ ranges, respectively, with deeper imaging providing less contamination and more high-redshift spectra.
That sample will thus fulfill the DESI requirements: with the expected fiber assignment rate of 0.69, it should provide more than 400 deg$^{-2}$ observed, reliable $\zspec$ in both the $0.6 < z < 1.1$ and the $1.1 < z < 1.6$ ranges.

% elg_vlo efficiency
The ELG\_VLO selection has an efficiency of 0 to 1, 67 to 70, and 21 to 26 percent in the $0 < z < 0.6$, $0.6 < z < 1.1$, and $1.1 < z < 1.6$ ranges, respectively.
As expected from its design, that sample has an overall very high efficiency, extremely few contaminants, and peaks in the $0.6 < z < 1.1$ range.
With the expected fiber assignment rate of 0.42, that should provide 130 deg$^{-2}$ and 50 deg$^{-2}$ reliable $\zspec$ in the $0.6 < z < 1.1$ and the $1.1 < z < 1.6$ ranges, respectively.\\

%=======================================================
% Acknowledgements
%=======================================================

% acknowledgements don t print out well with aastex631
% \begin{acknowledgments}
\vspace{\baselineskip}
\section*{Acknowledgments}

% JM
JM gratefully acknowledges support from the U.S. Department of Energy, Office of Science, Office of High Energy Physics under Award Number DE-SC0020086.

% desi
This research is supported by the Director, Office of Science, Office of High Energy Physics of the U.S. Department of Energy under Contract No. DE–AC02–05CH11231, and by the National Energy Research Scientific Computing Center, a DOE Office of Science User Facility under the same contract; additional support for DESI is provided by the U.S. National Science Foundation, Division of Astronomical Sciences under Contract No. AST-0950945 to the NSF’s National Optical-Infrared Astronomy Research Laboratory; the Science and Technologies Facilities Council of the United Kingdom; the Gordon and Betty Moore Foundation; the Heising-Simons Foundation; the French Alternative Energies and Atomic Energy Commission (CEA); the National Council of Science and Technology of Mexico (CONACYT); the Ministry of Science and Innovation of Spain (MICINN), and by the DESI Member Institutions: \niceurl{https://www.desi.lbl.gov/collaborating-institutions}.

% ls-dr9
The DESI Legacy Imaging Surveys consist of three individual and complementary projects: the Dark Energy Camera Legacy Survey (DECaLS), the Beijing-Arizona Sky Survey (BASS), and the Mayall z-band Legacy Survey (MzLS). DECaLS, BASS and MzLS together include data obtained, respectively, at the Blanco telescope, Cerro Tololo Inter-American Observatory, NSF’s NOIRLab; the Bok telescope, Steward Observatory, University of Arizona; and the Mayall telescope, Kitt Peak National Observatory, NOIRLab. NOIRLab is operated by the Association of Universities for Research in Astronomy (AURA) under a cooperative agreement with the National Science Foundation. Pipeline processing and analyses of the data were supported by NOIRLab and the Lawrence Berkeley National Laboratory. Legacy Surveys also uses data products from the Near-Earth Object Wide-field Infrared Survey Explorer (NEOWISE), a project of the Jet Propulsion Laboratory/California Institute of Technology, funded by the National Aeronautics and Space Administration. Legacy Surveys was supported by: the Director, Office of Science, Office of High Energy Physics of the U.S. Department of Energy; the National Energy Research Scientific Computing Center, a DOE Office of Science User Facility; the U.S. National Science Foundation, Division of Astronomical Sciences; the National Astronomical Observatories of China, the Chinese Academy of Sciences and the Chinese National Natural Science Foundation. LBNL is managed by the Regents of the University of California under contract to the U.S. Department of Energy. The complete acknowledgments can be found at \niceurl{https://www.legacysurvey.org}.

% kitt peak
The authors are honored to be permitted to conduct scientific research on Iolkam Du’ag (Kitt Peak), a mountain with particular significance to the Tohono O’odham Nation.

% \end{acknowledgments}

%=======================================================
% Data avaibility
%=======================================================

\vspace{\baselineskip}
\section*{Data avaibility}

The ELG targets for the SV1, the One-Percent, and the Main Surveys are accessible at \niceurl{https://data.desi.lbl.gov/public/ets/target/}.
We refer the reader to \citet{myers22a} for a description of the files and of the folders structure.

All data points shown in the published graphs are available in a machine-readable form on the following website: \niceurl{https://doi.org/10.5281/zenodo.6950999}.

%=======================================================
% Facility
%=======================================================

\facility{Mayall}

%=======================================================
% Bibliography
%=======================================================

\bibliography{ms}{}
\bibliographystyle{aasjournal}

\appendix
\numberwithin{table}{section}
\numberwithin{figure}{section}

%=======================================================
% Section : Extended (SV1) Target selection cuts
%=======================================================
\section{SV1 Survey Target selection cuts} \label{app:svcuts}

We present in the Table~\ref{tab:svcuts} the detailed cuts of the Survey Validation SV1 Survey ELG selection discussed in the Section~\ref{sec:svts}.
This selection is the union of two samples, SVGTOT and SVGFIB, which have similar cuts but are based either on $g_{\rm tot}$ or $g_{\rm fib}$. 
The overall selection has a target density of $\sim$7000 deg$^{-2}$, as the SVGFIB and SVGTOT selections have a large overlap.

The names SVGTOT and SVGFIB are names assigned to targeting bits by \texttt{desitarget}, the target selection pipeline \citep{myers22a}.
For completeness, we also report the cuts for the FDRGTOT and FDRGIB selections, which are other targeting bits.
The FDRGTOT (FDRGFIB, respectively) is fully included in SVGTOT (SVGFIB, respectively).\\

% Table: Extended (SV1) cuts
\begin{table*}
	\centering
	\begin{tabular}{lccl}
		\hline
		\hline
		Sample & Density & Cuts & Comment\\
		\hline
		% clean
		\multirow{4}{*}{Clean} & \multirow{4}{*}{-} & \texttt{brick\_primary} = True & Unique object\\
		& & $\texttt{nobs\_\{grz\}} > 0$ & Observed in the $grz$-bands\\
		& & $\texttt{flux\_\{grz\}} \times \sqrt{\texttt{flux\_ivar\_\{grz\}}} > 0$ & Positive SNR in the $grz$-bands\\
		& & $(\texttt{maskbits} \; \& \; 2^1) = 0$, $(\texttt{maskbits} \; \& \; 2^{12}) = 0$, $(\texttt{maskbits} \; \& \; 2^{13}) = 0$ & Not close to bright star/galaxy\\
		\hline
		% SVGTOT
		\multirow{5}{*}{SVGTOT} & \multirow{5}{*}{$\sim$5200 deg$^{-2}$} & Clean & Clean sample\\
		& & $g > 20$ & Bright cut\\
		& & $(g - r) + 1.2 \times (r - z) < 1.6 - 7.2 \times (g_{\rm tot} - \texttt{GTOTFAINT\_FDR})$ & Sliding faint cut\\
		& & $(g - r < 0.2)$ or $(g - r < 1.15 \times (r - z) + \texttt{LOWZCUT\_ZP} + 0.10)$ &Star/low-z cut\\
		& & $g - r < -1.2 \times (r - z) + 2.0$ & Redshift/\oii cut\\
        \hline
		% SVGFIB
		\multirow{5}{*}{SVGFIB} & \multirow{5}{*}{$\sim$5600 deg$^{-2}$} & Clean & Clean sample\\
		& & $g > 20$ & Bright cut\\
		& & $(g - r) + 1.2 \times (r - z) < 1.6 - 7.2 \times (g_{\rm fib} - \texttt{GFIBFAINT\_FDR})$ & Sliding faint cut\\
		& & $(g - r < 0.2)$ or $(g - r < 1.15 \times (r - z) + \texttt{LOWZCUT\_ZP} + 0.10)$ &Star/low-z cut\\
		& & $g - r < -1.2 \times (r - z) + 2.0$ & Redshift/\oii cut\\
        \hline
		% FDRGTOT
		\multirow{5}{*}{FDRGTOT} & \multirow{5}{*}{$\sim$2400 deg$^{-2}$} & Clean & Clean sample\\
		& & $g > 20$ & Bright cut\\
		& & $g < \texttt{GTOTFAINT\_FDR}$ & Faint cut\\
		& & $0.3 < r - z < 1.6$ & $r - z$ cut\\
		& & $g - r < 1.15 \times (r - z) + \texttt{LOWZCUT\_ZP}$ &Star/low-z cut\\
		& & $g - r < -1.2 \times (r - z) + 1.6$ & Redshift/\oii cut\\
        \hline
		% FDRGFIB
		\multirow{5}{*}{FDRGFIB} & \multirow{5}{*}{$\sim$2500 deg$^{-2}$} & Clean & Clean sample\\
		& & $g > 20$ & Bright cut\\
		& & $g_{\rm fib} < \texttt{GFIBFAINT\_FDR}$ & Faint cut\\
		& & $0.3 < r - z < 1.6$ & $r - z$ cut\\
		& & $g - r < 1.15 \times (r - z) + \texttt{LOWZCUT\_ZP}$ &Star/low-z cut\\
		& & $g - r < -1.2 \times (r - z) + 1.6$ & Redshift/\oii cut\\
        \hline
    \end{tabular}
	\caption{
		Survey Validation SV1 Survey target selection cuts.
		The following quantities have values defined for the North and the South region:
	    $\texttt{GTOTFAINT\_FDR}: \rm{North}=23.5, \; \rm{South}=23.4$;
	    $\texttt{GFIBFAINT\_FDR}: \rm{North}=24.1, \; \rm{South}=24.1$;
	    $\texttt{LOWZCUT\_ZP}: \rm{North}=-0.20, \; \rm{South}=-0.15$.
        See Table~\ref{tab:maincuts} for the definitions of the terms in the cuts.
	}
	\label{tab:svcuts}
\end{table*}

%=======================================================
% Section : HSC/DR2 zphot vs. DESI/ELG zspec
%=======================================================
\section{HSC/DR2 $\zphot$ comparison to DESI/ELG $\zspec$} \label{app:hsczphot}

We illustrate in Figure~\ref{fig:desi_hsc} how the HSC/DR2 $\zphot$ estimates perform against the Main Survey ELG DESI $\zspec$ measurements corresponding to observations until December 2021 (see Section~\ref{sec:specobs}).
We emphasize that DESI $\zspec$ data were not used to train the HSC/DR2 $\zphot$ method.
The matched sample has $\sim$175 thousand targets with a $\zspec$ passing our reliability criterion of Equation~\ref{eq:zcrit}.
We do not make any quality cuts on the HSC $\zphot$.

The left-hand panels compare the two redshift measurements, the color-coding indicating the HSC \texttt{risk} parameter, which quantifies the reliability of the $\zphot$ estimate \citep[with $\texttt{risk}  = 0$ being the most secure and $\texttt{risk} = 1$ the less secure; see][]{tanaka18a}.
Overall the HSC $\zphot$ estimates perform very well, noticeably over the whole $0.6 < z < 1.6$ range, thanks to its very deep imaging and the presence of $y$-band imaging.
This justifies the use of the HSC $\zphot$ in Figures~\ref{fig:maingrz} and \ref{fig:svgrz}.
In particular, the \texttt{risk} parameter provides a sensible estimation of the $\zphot$ reliability.
This last point is illustrated in the bottom panel of the left-hand plots, where we also display the median value of $\Delta z = (\zphot - \zspec) / (1 + \zspec)$ for three subsamples: all matches (black), the 50 percent lower \texttt{risk} parameter values (cyan), and the 25 percent lower \texttt{risk} parameter values (red).
The $1.48 \times \rm MAD(\Delta z)$ values are displayed as shaded regions, with typical values of 0.08 (all matches), 0.05 (50 percent lower \texttt{risk}), and 0.02 (25 percent lower \texttt{risk}).

Nevertheless, those cuts on \texttt{risk} are biasing the sample towards the redder ELG targets -- hence the lower redshift ones, as those are easier to model by the $\zphot$ algorithms; said differently, applying cuts on \texttt{risk} will exclude from the sample the blue, high-redshift ELGs.
That is illustrated in the right-hand panels, where we plot for each subsample (all, 50 percent and 25 percent lower \texttt{risk} parameter values) the DESI $\zspec$ distribution (filled histograms) and the HSC $\zphot$ distribution (empty histograms); the DESI $\zspec$ distribution with no cut on \texttt{risk} is displayed with a black, hatched histogram.
As a consequence, a careful balance between the $\zphot$ reliability and the sample representativeness will be required for any analysis using the HSC $\zphot$ distribution for an ELG DESI-like sample, as for instance trying to infer the redshift distribution of the DESI ELG targets which do not provide a reliable $\zspec$.

% figure: desi vs. hsc
\begin{figure*}[!h]
	\begin{center}
		\begin{tabular}{lr}
			\includegraphics[height=6cm,keepaspectratio]{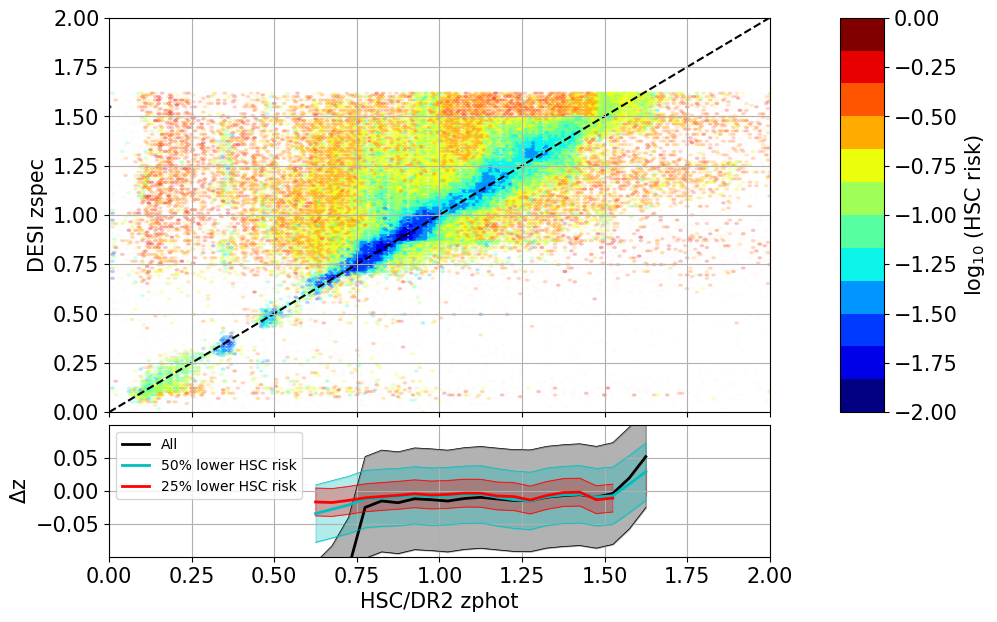} & 
			\includegraphics[height=6cm,keepaspectratio]{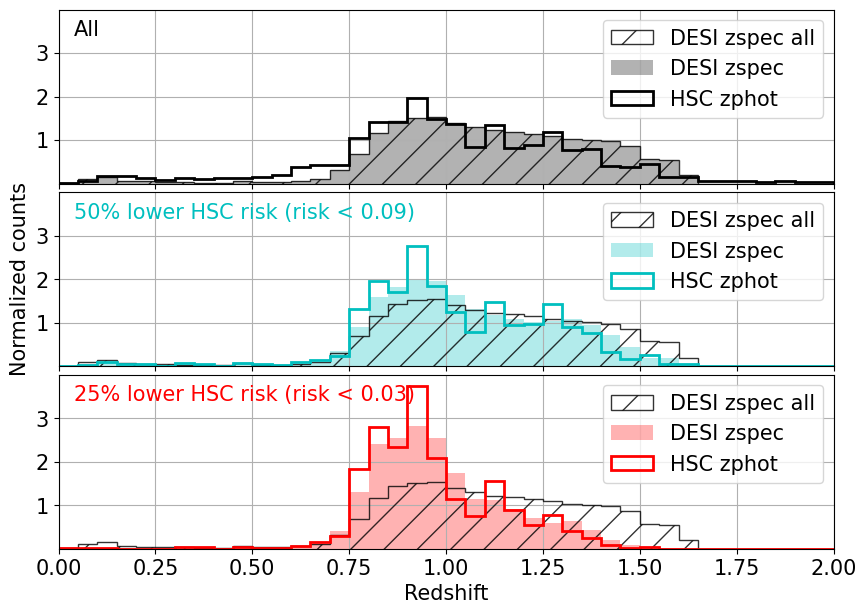}\\
		\end{tabular}
		\caption{
			Comparison of the HSC/DR2 $\zphot$ to the DESI Main Survey ELG $\zspec$ measurements, using $\sim$175 thousand matched targets with a $\zspec$ passing our reliability criterion of Equation~\ref{eq:zcrit}.
			\textit{Lef-hand panels}: scatter plots color-coded according to the average HSC \texttt{risk} parameter (see text); the solid lines in the bottom panel are the median values of $\Delta z = (\zphot - \zspec) / (1 + \zspec)$ for three subsamples: all the matches (black), the 50 percent lower \texttt{risk} parameter values (cyan), and the 25 percent lower \texttt{risk} parameter values (red); the shaded regions show the $1.48 \times \rm MAD(\Delta z)$ values.
			\textit{Right-hand panels}: redshift distributions for those three subsamples; in each panel, the filled (empty, respectively) histogram is for the DESI $\zspec$ (HSC $\zphot$, respectively).
			The DESI $\zspec$ distribution with no cut on \texttt{risk} is displayed with a black, hatched histogram.
		}
		\label{fig:desi_hsc}
	\end{center}
\end{figure*}

\end{document}

%% file: DESI-2021-0104_author_list.tex
% Author list file generated with: mkauthlist 1.2.4+25.gdac5248 
% mkauthlist DESI-2021-0104_author_list.csv DESI-2021-0104_author_list.tex -j aj -f --sort --orcid 

\author[0000-0001-5999-7923]{A.~Raichoor}
\affiliation{Lawrence Berkeley National Laboratory, 1 Cyclotron Road, Berkeley, CA 94720, USA}

\author[0000-0002-2733-4559]{J.~Moustakas}
\affiliation{Department of Physics and Astronomy, Siena College, 515 Loudon Road, Loudonville, NY 12211, USA}

\author[0000-0001-8684-2222]{Jeffrey A.~Newman}
\affiliation{Department of Physics \& Astronomy and Pittsburgh Particle Physics, Astrophysics, and Cosmology Center (PITT PACC), University of Pittsburgh, 3941 O'Hara Street, Pittsburgh, PA 15260, USA}

\author[0000-0002-5652-8870]{T.~Karim}
\affiliation{Center for Astrophysics $|$ Harvard \& Smithsonian, 60 Garden Street, Cambridge, MA 02138, USA}

\author[0000-0001-6098-7247]{S.~Ahlen}
\affiliation{Physics Dept., Boston University, 590 Commonwealth Avenue, Boston, MA 02215, USA}

\author[0000-0002-3757-6359]{Shadab Alam}
\affiliation{Institute for Astronomy, University of Edinburgh, Royal Observatory, Blackford Hill, Edinburgh EH9 3HJ, UK}

\author[0000-0003-4162-6619]{S.~Bailey}
\affiliation{Lawrence Berkeley National Laboratory, 1 Cyclotron Road, Berkeley, CA 94720, USA}

\author{D.~Brooks}
\affiliation{Department of Physics \& Astronomy, University College London, Gower Street, London, WC1E 6BT, UK}

\author{K.~Dawson}
\affiliation{Department of Physics and Astronomy, The University of Utah, 115 South 1400 East, Salt Lake City, UT 84112, USA}

\author{A.~de la Macorra}
\affiliation{Instituto de F\'{\i}sica, Universidad Nacional Aut\'{o}noma de M\'{e}xico,  Cd. de M\'{e}xico  C.P. 04510,  M\'{e}xico}

\author{A.~de~Mattia}
\affiliation{IRFU, CEA, Universit\'{e} Paris-Saclay, F-91191 Gif-sur-Yvette, France}

\author[0000-0002-4928-4003]{A.~Dey}
\affiliation{NSF's NOIRLab, 950 N. Cherry Ave., Tucson, AZ 85719, USA}

\author[0000-0002-5665-7912]{Biprateep~Dey}
\affiliation{Department of Physics \& Astronomy and Pittsburgh Particle Physics, Astrophysics, and Cosmology Center (PITT PACC), University of Pittsburgh, 3941 O'Hara Street, Pittsburgh, PA 15260, USA}

\author[0000-0002-5402-1216]{G.~Dhungana}
\affiliation{Department of Physics, Southern Methodist University, 3215 Daniel Avenue, Dallas, TX 75275, USA}

\author{S.~Eftekharzadeh}
\affiliation{Universities Space Research Association, NASA Ames Research Centre}

\author{D.~J.~Eisenstein}
\affiliation{Center for Astrophysics $|$ Harvard \& Smithsonian, 60 Garden Street, Cambridge, MA 02138, USA}

\author{K.~Fanning}
\affiliation{Department of Physics, The Ohio State University, 191 West Woodruff Avenue, Columbus, OH 43210, USA}
\affiliation{Center for Cosmology and AstroParticle Physics, The Ohio State University, 191 West Woodruff Avenue, Columbus, OH 43210, USA}

\author[0000-0002-3033-7312]{A.~Font-Ribera}
\affiliation{Institut de F\'{i}sica d’Altes Energies (IFAE), The Barcelona Institute of Science and Technology, Campus UAB, 08193 Bellaterra Barcelona, Spain}

\author{J.~Garc\'ia-Bellido}
\affiliation{Instituto de F\'{\i}sica Te\'{o}rica (IFT) UAM/CSIC, Universidad Aut\'{o}noma de Madrid, Cantoblanco, E-28049, Madrid, Spain}

\author{E.~Gaztañaga}
\affiliation{Institut d'Estudis Espacials de Catalunya (IEEC), 08034 Barcelona, Spain}
\affiliation{Institute of Space Sciences, ICE-CSIC, Campus UAB, Carrer de Can Magrans s/n, 08913 Bellaterra, Barcelona, Spain}

\author{S.~Gontcho A Gontcho}
\affiliation{Lawrence Berkeley National Laboratory, 1 Cyclotron Road, Berkeley, CA 94720, USA}

\author{J.~Guy}
\affiliation{Lawrence Berkeley National Laboratory, 1 Cyclotron Road, Berkeley, CA 94720, USA}

\author{K.~Honscheid}
\affiliation{Center for Cosmology and AstroParticle Physics, The Ohio State University, 191 West Woodruff Avenue, Columbus, OH 43210, USA}
\affiliation{Department of Physics, The Ohio State University, 191 West Woodruff Avenue, Columbus, OH 43210, USA}

\author[0000-0002-6024-466X]{M.~Ishak}
\affiliation{Department of Physics, The University of Texas at Dallas, Richardson, TX 75080, USA}

\author{R.~Kehoe}
\affiliation{Department of Physics, Southern Methodist University, 3215 Daniel Avenue, Dallas, TX 75275, USA}

\author[0000-0003-3510-7134]{T.~Kisner}
\affiliation{Lawrence Berkeley National Laboratory, 1 Cyclotron Road, Berkeley, CA 94720, USA}

\author[0000-0001-6356-7424]{A.~Kremin}
\affiliation{Lawrence Berkeley National Laboratory, 1 Cyclotron Road, Berkeley, CA 94720, USA}

\author[0000-0001-8857-7020]{Ting-Wen Lan}
\affiliation{Graduate Institute of Astrophysics and Department of Physics, National Taiwan University, No. 1, Sec. 4, Roosevelt Rd., Taipei 10617, Taiwan}

\author[0000-0003-1838-8528]{M.~Landriau}
\affiliation{Lawrence Berkeley National Laboratory, 1 Cyclotron Road, Berkeley, CA 94720, USA}

\author[0000-0001-7178-8868]{L.~Le~Guillou}
\affiliation{Sorbonne Universit\'{e}, CNRS/IN2P3, Laboratoire de Physique Nucl\'{e}aire et de Hautes Energies (LPNHE), FR-75005 Paris, France}

\author[0000-0003-1887-1018]{Michael E.~Levi}
\affiliation{Lawrence Berkeley National Laboratory, 1 Cyclotron Road, Berkeley, CA 94720, USA}

\author{C.~Magneville}
\affiliation{IRFU, CEA, Universit\'{e} Paris-Saclay, F-91191 Gif-sur-Yvette, France}

\author[0000-0002-4279-4182]{P.~Martini}
\affiliation{Center for Cosmology and AstroParticle Physics, The Ohio State University, 191 West Woodruff Avenue, Columbus, OH 43210, USA}
\affiliation{Department of Astronomy, The Ohio State University, 4055 McPherson Laboratory, 140 W 18th Avenue, Columbus, OH 43210, USA}

\author[0000-0002-1125-7384]{Aaron M. Meisner}
\affiliation{NSF's NOIRLab, 950 N. Cherry Ave., Tucson, AZ 85719, USA}

\author{Adam~D.~Myers}
\affiliation{Department of Physics \& Astronomy, University  of Wyoming, 1000 E. University, Dept.~3905, Laramie, WY 82071, USA}

\author[0000-0001-6590-8122]{Jundan Nie}
\affiliation{National Astronomical Observatories, Chinese Academy of Sciences, A20 Datun Rd., Chaoyang District, Beijing, 100012, P.R. China}

\author[0000-0003-3188-784X]{N.~Palanque-Delabrouille}
\affiliation{Lawrence Berkeley National Laboratory, 1 Cyclotron Road, Berkeley, CA 94720, USA}
\affiliation{IRFU, CEA, Universit\'{e} Paris-Saclay, F-91191 Gif-sur-Yvette, France}

\author[0000-0002-0644-5727]{W.J.~Percival}
\affiliation{Department of Physics and Astronomy, University of Waterloo, 200 University Ave W, Waterloo, ON N2L 3G1, Canada}
\affiliation{Perimeter Institute for Theoretical Physics, 31 Caroline St. North, Waterloo, ON N2L 2Y5, Canada}
\affiliation{Waterloo Centre for Astrophysics, University of Waterloo, 200 University Ave W, Waterloo, ON N2L 3G1, Canada}

\author{C.~Poppett}
\affiliation{Lawrence Berkeley National Laboratory, 1 Cyclotron Road, Berkeley, CA 94720, USA}
\affiliation{Space Sciences Laboratory, University of California, Berkeley, 7 Gauss Way, Berkeley, CA  94720, USA}
\affiliation{University of California, Berkeley, 110 Sproul Hall \#5800 Berkeley, CA 94720, USA}

\author[0000-0001-7145-8674]{F.~Prada}
\affiliation{Instituto de Astrof\'{i}sica de Andaluc\'{i}a (CSIC), Glorieta de la Astronom\'{i}a, s/n, E-18008 Granada, Spain}

\author{A.~J.~Ross}
\affiliation{Center for Cosmology and AstroParticle Physics, The Ohio State University, 191 West Woodruff Avenue, Columbus, OH 43210, USA}

\author{V.~Ruhlmann-Kleider}
\affiliation{IRFU, CEA, Universit\'{e} Paris-Saclay, F-91191 Gif-sur-Yvette, France}

\author[0000-0002-5513-5303]{C.~G.~Sabiu}
\affiliation{Natural Science Research Institute, University of Seoul, 163 Seoulsiripdae-ro, Dongdaemun-gu, Seoul, South Korea}

\author[0000-0002-3569-7421]{E.~F.~Schlafly}
\affiliation{Space Telescope Science Institute, 3700 San Martin Drive, Baltimore, MD 21218, USA}

\author{D.~Schlegel}
\affiliation{Lawrence Berkeley National Laboratory, 1 Cyclotron Road, Berkeley, CA 94720, USA}

\author[0000-0003-1704-0781]{Gregory~Tarl\'{e}}
\affiliation{University of Michigan, Ann Arbor, MI 48109, USA}

\author{B.~A.~Weaver}
\affiliation{NSF's NOIRLab, 950 N. Cherry Ave., Tucson, AZ 85719, USA}

\author[0000-0001-5146-8533]{Christophe~Yèche}
\affiliation{IRFU, CEA, Universit\'{e} Paris-Saclay, F-91191 Gif-sur-Yvette, France}

\author[0000-0001-5381-4372]{Rongpu Zhou}
\affiliation{Lawrence Berkeley National Laboratory, 1 Cyclotron Road, Berkeley, CA 94720, USA}

\author[0000-0002-4135-0977]{Zhimin~Zhou}
\affiliation{National Astronomical Observatories, Chinese Academy of Sciences, A20 Datun Rd., Chaoyang District, Beijing, 100012, P.R. China}

\author[0000-0002-6684-3997]{H.~Zou}
\affiliation{National Astronomical Observatories, Chinese Academy of Sciences, A20 Datun Rd., Chaoyang District, Beijing, 100012, P.R. China}